\documentclass[aps, prx, twocolumn, superscriptaddress, amsmath, amssymb, floatfix,longbibliography, notitlepage]{revtex4-2}
\usepackage{amsbsy}
\usepackage{amsfonts}
\usepackage{amsmath}
\usepackage{amstext}
\usepackage{amsthm}
\usepackage{amssymb}
\usepackage{braket}
\usepackage{xparse}
\usepackage{enumitem}

\usepackage{graphicx}

\usepackage{epsfig}
\usepackage{epstopdf}
\usepackage{bm}
\usepackage{color}
\usepackage[dvipsnames]{xcolor}
\definecolor{prlblue}{rgb}{0.18,0.19,0.57}
\usepackage{multirow}
\usepackage[colorlinks]{hyperref}
\usepackage{cleveref}

\usepackage{xparse}

\makeatletter
\NewDocumentCommand{\sump}{e{_}}
 {%
  \DOTSB
  \mathop{\IfNoValueTF{#1}{\sump@{}}{\sump@{#1}}}%
  \nolimits
 }
\newcommand{\sump@}[1]{\mathpalette\sump@@{#1}}
\newcommand{\sump@@}[2]{%
  \ifx#1\displaystyle
    {\sump@display{#2}}%
  \else
    \sum@\nolimits'_{#2}%
  \fi
}
\newcommand{\sump@display}[1]{%
  \sbox\z@{$\m@th\displaystyle\sum@\nolimits'$}%
  \sbox\tw@{$\m@th\displaystyle\sum@\limits_{#1}$}%
  \sbox\@tempboxa{$\m@th\displaystyle'$}
  \mathop{\sum@\nolimits' \kern-\wd\@tempboxa}\limits_{#1}%
  \ifdim\wd\z@>\wd\tw@
    \kern\dimexpr\wd\z@-\wd\tw@\relax
  \fi
}
\makeatother

\hypersetup{
    colorlinks=true,
    citecolor=Green,
    linkcolor=Red,
    urlcolor=NavyBlue
}

\newcommand{\diagram}[2]{\vcenter{\hbox{\includegraphics[height=#2]{#1}}}}

\begin{document}
\title{Momentum-Selective Two-Component Excitations in Electron-Doped Mott Insulators
}


\author{Zeyu Han}
\thanks{These two authors contributed equally to this work}
\affiliation{Institute for Advanced Study, Tsinghua University, Beijing 100084, China}

\author{Can Cui}
\thanks{These two authors contributed equally to this work}
\affiliation{Institute for Advanced Study, Tsinghua University, Beijing 100084, China}

\author{Jia-Xin Zhang}
\email{jiaxin\_zhang@ucsb.edu}
\affiliation{French American Center for Theoretical Science, CNRS, KITP, Santa Barbara, California 93106-4030, USA}
\affiliation{Kavli Institute for Theoretical Physics, University of California, Santa Barbara, California 93106-4030, USA}

\author{Zheng-Yu Weng}
\affiliation{Institute for Advanced Study, Tsinghua University, Beijing 100084, China}

\date{\today}

\begin{abstract}
The asymmetry in low-energy single-particle excitations between electron-doped and hole-doped cuprates has been extensively examined experimentally. Electron-doped cuprates exhibit a nontrivial dichotomy, in which largely Fermi-liquid-like behavior, suggestive of comparatively weak electronic correlations, coexists with correlation-driven features reminiscent of hole-doped systems, thereby posing a significant challenge to a unified understanding of the underlying physics within a doped-Mott-insulator framework.
The present work addresses this issue by first establishing that, within the $t$-$t'$-$J$ model, the ground-state wave function 
generically admits a two-component structure consisting of a coherent quasiparticle component and an incoherent composite component.
The ground-state kinetic energy can be understood as arising from both the intrinsic propagation of the coherent quasiparticle and a resonance between these two components. Based on our variational Monte Carlo results at the single-hole level, we find that on the hole-doped side ($t'<0$), intercomponent resonance becomes the dominant contribution and is concentrated in the low-energy nodal region. This resonance-induced form of emergent single-particle propagation can be physically interpreted as originating from the recombination of fractionalized degrees of freedom, which underlies a variety of unconventional phenomena driven by strong correlations. On the other hand, on the electron-doped side ($t'>0$), the propagation of the coherent quasiparticle component, which carries a more conventional Fermi-liquid-like character, is selectively enhanced in the low-energy antinodal region. This naturally leads to a two-component description that is well separated in momentum space on the electron-doped side, where the low-energy antinodal spectral weight is dominated by the coherent quasiparticle and is fundamentally distinct from that in the nodal region, which remains dominated by the incoherent composite component. Motivated by such a structure and guided by experimental observations, we propose a phenomenological Green's function at finite doping, yielding spectral features consistent with experiments.

\end{abstract}

\maketitle

\tableofcontents

\section{Introduction}
\label{Introduction}
The marked asymmetry between hole- and electron-doped cuprates remains poorly understood and continues to challenge our understanding of high-temperature superconductivity. In particular, one of the most prominent asymmetries is evident in the low-energy single-particle spectral weight.

In hole-doped cuprates, angle-resolved photoemission spectroscopy (ARPES) experiments reveal that, in the normal state, the low-energy single-particle spectral weight does not form a closed Fermi surface, but instead appears as disconnected segments known as Fermi arcs centered around the nodal regions
~\cite{Shen_arpes_reviev,Ding1996,Norman1998,Shen_remnant_FS}. The Fermi arcs grow with doping, and eventually a full Fermi surface is restored on the overdoped side~\cite{Chatterjee_Full_FS,Kaminski_Full_FS}. In contrast to hole-doped cuprates, electron-doped cuprates exhibit seemingly more conventional behavior in spectroscopic measurements. In lightly doped samples, low-energy single-particle excitations are predominantly located in the antinodal region of the first Brillouin zone, forming electron pockets~\cite{Matsui2007,Armitage2002}, in sharp contrast to the behavior of hole-doped cuprates. Upon increasing doping, a small hole pocket gradually emerges in the nodal region~\cite{ArmitageRMP}, symmetrically positioned with respect to the antiferromagnetic zone boundary (AFMZB).
Finally, at higher doping, the antinodal electron pocket merges with the nodal pocket, eventually forming a large Fermi surface~\cite{KeJunXu,Gossamer_FS,Armitage2002,Tang2021}.

In addition to the asymmetry of the low-energy spectral weight, long-range antiferromagnetic (AFM) fluctuations observed in electron-doped cuprates are more robust than in their hole-doped counterparts~\cite{Motoyama2007,Lee2014}. This observation has motivated theoretical descriptions in which electrons are treated as itinerant coherent quasiparticles interacting with AFM fluctuations~\cite{Park_folding}. Although this framework captures certain features of the single-particle spectrum, its validity as a weak-coupling description warrants careful scrutiny by resolving the following issues. First, from a theoretical standpoint, if one accepts that the essential physics of hole-doped cuprates is rooted in the doped-Mott-insulator paradigm, it is natural to question why such strong-coupling physics would be replaced by a weak-coupling picture on the electron-doped side, particularly given their close proximity in the phase diagram. 
Second, on the experimental side, a variety of non–Fermi-liquid behaviors, generally weaker but qualitatively similar to those in hole-doped cuprates, have also been observed in electron-doped compounds, including the sign reversal of the Hall coefficient~\cite{Transport_FSR,Seebeck}, the sign change of the Seebeck coefficient~\cite{Seebeck,Mandal2019}, linear-in-field magnetoresistance~\cite{Jin2011}, antinodal kink~\cite{Tang2022npj}, linear-in-temperature resistivity in the strange-metal regime~\cite{Jin2011,annurev_Greene}. Furthermore, neutron-scattering experiments reveal a well-defined magnetic resonance mode characterized by an energy scale $E_g$, which scales with the superconducting transition temperature $T_c$ according to $E_g\approx 6T_c$~\cite{Wilson2006,NCCO_Eg,Yu2009}, with a similar energy scale also identified by Raman scattering~\cite{Stadlober_Eg}. This scaling is consistent with that observed in various hole-doped cuprate systems~\cite{Y123_Eg,Bi2212_Eg,Tl2201_Eg,La214_Eg,Bi2212_Eg1}. Altogether, these observations point toward a common underlying mechanism shared by hole- and electron-doped cuprates, thereby motivating a unified theoretical description and highlighting the limitations of weak-coupling approaches.

The single-band $t$-$t^\prime$-$J$ model~\cite{DopeMottInsulator,ArmitageRMP} is a promising microscopic model for capturing a wide range of experimentally observed phenomena. In particular, accumulating numerical evidence~\cite{Senechal2004,Moritz2009,DNSheng,Jiang_Shengtao,Zhao_Weng_2024,wang_wen_2025} indicates that the single-band $t$-$t^\prime$-$J$ model (and the closely related single-band $t$-$t^\prime$-$U$ model) exhibits single-particle spectral features that are qualitatively consistent with experimental observations. In density-matrix renormalization group (DMRG) studies of the ground-state correlations of the $t$-$t^\prime$-$J$ model \cite{Jiang_Shengtao}, it has been demonstrated that $t'>0$ significantly enhances antiferromagnetic correlations compared to the $t'<0$ case, consistent with the stronger AFM tendencies on the electron-doped side. Focusing on the spectral function, complementary variational Monte Carlo (VMC) studies of the single-hole–doped $t$–$t'$–$J$ model \cite{Zhao_Weng_2024} revealed a systematic transfer of low-energy spectral weight from the nodal to the antinodal region as $t'$ is tuned from negative to positive, in qualitative agreement with ARPES observations~\cite{Armitage2002,ShenZX_pnas,Shen_arpes_reviev}. More recently, determinant quantum Monte Carlo (DQMC) studies of the $t$–$t^\prime$–$U$ model \cite{wang_wen_2025} have revealed a similar nodal–antinodal dichotomy in the single-particle spectral function. This dichotomy is attributed to a strongly momentum-dependent damping rate, thereby leading to low-energy spectral weight concentrated in the nodal (antinodal) region on the hole- (electron-) doped side.

Taken together, these numerical approaches establish that, within the $t$-$J$ model itself, strong-correlation physics is already highly nontrivial, and that the next-nearest-neighbor (NNN) hopping $t^\prime$ plays a key role in producing the pronounced electron–hole asymmetry consistent with experimental observations~\cite{ArmitageRMP,Shen_arpes_reviev,DopeMottInsulator}. This fact motivates a systematic investigation of the ground-state structure of the $t$–$J$ model, and of how $t^\prime$ modifies it, in search of potential connections to cuprate superconductivity.

\begin{figure*}[!ht]
\centering
\includegraphics[width=1\linewidth]{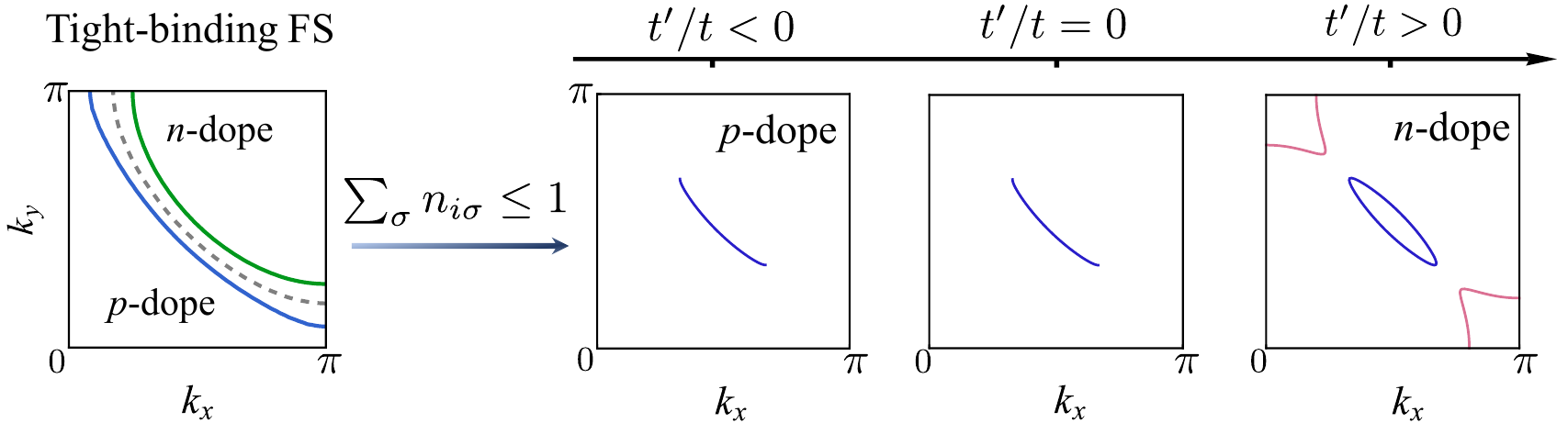}
\caption{Distinct evolution of Fermi surfaces (FS) away from half-filling ($\delta = 0$). (Left) Non-interacting tight-binding model with a fixed $t^\prime=-0.25t$ ($t>0$) in the first quadrant of the Brillouin zone, showing the Fermi surfaces for hole doping ($\delta = 0.15$, $p$-dope, blue line), electron doping ($\delta = 0.15$, $n$-dope, green line), and half-filling (gray dashed line) obtained by simply tuning the chemical potential. (Right) Schematic Fermi surfaces in the presence of a large Mott gap enforced by the no-double-occupancy constraint $\sum_{\sigma} n_{i\sigma} \leq 1$. Under this constraint, the electron-doped regime is mapped to a hole-doped model with a reversed sign of $t^\prime$. Therefore, the evolution is plotted along a parameter axis of the next-nearest-neighbor hopping $t^\prime/t$. For the hole-doped case corresponding to $t^\prime/t \le 0$, the low-energy spectral weight is concentrated near the nodal region, forming distinct Fermi arcs (blue lines). In contrast, for the electron-doped case mapped to the $t^\prime/t > 0$ regime, both the nodal pockets (blue ellipse) and antinodal regions (pink segments) contribute to the low-energy spectral weight. This highlights the asymmetric Fermi surface reconstructions between the electron- and hole-doped regimes under strong correlation.}
\label{Phase diagram}
\end{figure*}

In light of the experimental and numerical studies reviewed above, two key questions emerge naturally:
\begin{enumerate}[label=(\alph*),itemsep=0pt,parsep=0pt,topsep=0pt]
\item As illustrated in the left panel of Fig.~\ref{Phase diagram}, the non-interacting tight-binding Fermi surfaces exhibit only minor deformations when tuning the chemical potential for electron and hole doping. However, as long as the on-site Coulomb repulsion constrains the local Hilbert space with $\sum_\sigma n_{i\sigma} \le 1$, the original tight-binding models in the electron- and hole-doped regimes flow to a low-energy Hamiltonian with opposite signs of $t^\prime$. In the purely non-interacting limit, systems with opposite signs of $t^\prime$ are trivially related by a $(\pi,\pi)$ momentum shift in the Fermi surface, which is a direct consequence of the particle–hole transformation $c^\dagger_{i\sigma} \rightarrow (-1)^i c_{i\sigma}$. How, then, does the strong correlation constraint drive the dramatic reconstruction sketched in the right panels of Fig.~\ref{Phase diagram}, especially at $t^\prime/t>0$, shifting the low-energy excitations from the nodal to the antinodal region?
\item Building on the understanding of the previous question, can the pronounced electron–hole asymmetry observed at finite doping be systematically interpreted by disentangling the role of $t^\prime$?
\end{enumerate}

Recently, progress has been made in understanding the low-energy physics of the single-hole-doped $t$–$J$ model~\cite{cancui}, in which the low-lying wave functions are decomposed into a linear superposition of a coherent quasiparticle component and an incoherent one.
Within this framework, the propagation of a doped hole is fundamentally distinct from that of a Landau quasiparticle, for which the incoherent background merely manifests as a passive backflow and only renormalizes the effective mass. Instead, the kinetic energy of the doped hole is gained predominantly through tunneling processes between the coherent and incoherent components. This formulation therefore provides a controlled and physically transparent setting for investigating the effects of the NNN hopping $t^\prime$.

In this work, to tackle question (a), we perform VMC simulations of the single- and two-hole-doped $t$–$t^\prime$–$J$ model. Our findings reveal that in the hole-doped regime ($t^\prime<0$), the dynamics of doped holes are fundamentally inherited from the $t^\prime=0$ limit. In this regime, a bare hole gains kinetic energy mainly via \emph{resonance} between the coherent and incoherent sectors. This mechanism yields emergent low-energy coherent quasiparticle excitations residing near the nodal region. Conversely, in the electron-doped regime ($t^\prime>0$), beyond a threshold value of $t^\prime/t=0.09$, a stark enhancement of the coherent quasiparticle weight develops at low energies in the antinodal region, driven by the intrinsic propagation of bare electrons. Meanwhile, the resonance-induced nodal excitations persist largely unaltered, save for a minor energy shift. Turning to the two-hole sector, the ground state for both signs of $t^{\prime}$ exhibits a tightly bound pairing structure consisting of a nearest-neighbor (NN) $d_{x^2-y^2}$ Cooper-pairing component coexisting with an incoherent NNN counterpart. Notably, a positive $t^{\prime}$ facilitates the intrinsic propagation of Cooper pairs through the NNN hopping channel. Consequently, the low-lying spectral weight in momentum space is redirected from the nodal lines ($k_x = \pm k_y$) toward the antinodal region.

These results reveal a dichotomy at finite doping, manifesting not only between the electron- and hole-doped regimes, but also within the momentum space of the electron-doped side. Although this nodal–antinodal dichotomy has previously been proposed in numerical studies~\cite{wang_wen_2025} and inferred from experimental observations~\cite{ShenZX_pnas,tang2022} through close examination of the single-particle damping rate, our VMC results further clarify its underlying physical mechanism and establish a direct connection to the strong-correlation effects already present in the $t$–$J$ model without $t'$.

Our numerical calculations, although performed in the extremely low-doping limit, nevertheless provide hints toward an answer to the second question we raised, enabling us to gain insight into the finite-doping regime. In addition, we present a discussion of the $t$–$t^\prime$–$J$ model at finite doping, focusing primarily on the structure of the single-particle spectrum. A phenomenological single-particle Green’s function is constructed, guided by the constraints imposed by our numerical results and by the following experimental observations:
  \begin{enumerate}[label=(\roman*),itemsep=0pt,parsep=0pt,topsep=0pt]
    \item Long-range AFM fluctuations persist over a wider doping range in electron-doped cuprates compared to their hole-doped counterparts~\cite{Motoyama2007}.
    \item Single-particle spectra of nearly half-filled cuprates point toward a Mott-insulating ground state rather than a weak-coupling Slater insulator~\cite{Halffilled}.
    \item Despite systematically smaller superconducting gap scales in the electron-doped regime than in the hole-doped regime~\cite{Niestemski2007}, the ratio $2\Delta/k_{\mathrm{B}}T_c$ has been reported to reach $\approx 7.5$, suggesting a strong-coupling origin for the superconductivity.
    \item A well-defined magnetic resonance mode is observed in electron-doped cuprates~\cite{Stadlober_Eg,Wilson2006}. The resonance energy $E_g$ scales with the superconducting transition temperature $T_c$ approximately as $E_g \approx 6T_c$, similar to the scaling observed in the hole-doped regime~\cite{Y123_Eg,Bi2212_Eg,Tl2201_Eg,La214_Eg,Bi2212_Eg1}.
    \item At finite doping, a finite spectral weight emerges in the nodal region, irrespective of whether the system is electron- or hole-doped~\cite{ArmitageRMP,Armitage2002,Shen_arpes_reviev,Norman1998}.
    \item Experimental measurements strongly suggest a nodal–antinodal dichotomy in electron-doped cuprates, namely that single-particle spectra exhibit qualitatively distinct behaviors in the nodal and antinodal regions~\cite{ShenZX_pnas,tang2022}.
\end{enumerate}
These compelling experimental facts motivate the present study from three key perspectives. First, points (i) through (iii) accentuate the indispensable role of strong correlations, necessitating a meticulous treatment of AFM fluctuations. Second, observations (iv) and (v) reveal that superconductivity and the low-energy nodal spectral weight share a unified physical origin in both doping regimes, suggesting that the finite-doping physics of electron- and hole-doped cuprates is deeply connected. Third, point (vi) establishes the presence of a robust nodal–antinodal dichotomy in momentum space at finite doping. The comparatively weaker non-Fermi-liquid features on the electron-doped side can thus be attributed to the advent of an additional, less correlated quasiparticle mode.

More specifically, building upon the phase-string formulation of the $t$–$J$ model~\cite{Weng1997,Weng1999,Ma_2014}, we propose a two-component description for the $t^\prime>0$ regime. This framework is starkly distinct from conventional two-component pictures encountered in multi-orbital systems, such as heavy-fermion materials where localized $d$- or $f$-electron moments hybridize with itinerant conduction electrons~\cite{HeavyFermion}, as well as nickelates and iron-based superconductors~\cite{NdNiO2,YuRong,OSMT,Kou_2009}. In these conventional systems, the coexistence of multiple components arises explicitly from orbital degrees of freedom~\cite{You_2014}. In contrast, the two-component structure identified here is purely correlation-driven rather than rooted in orbital differentiation. Instead, the two coexisting fluids emerge from distinct momentum sectors of the Brillouin zone. Consequently, the resulting single-particle excitations acquire qualitatively different characteristics depending on their locations in momentum space, as will be demonstrated by the detailed spectral analysis presented in the following sections.

The remainder of the paper is organized as follows. In Sec.~\ref{RevisitPS}, we briefly review the sign structure of the $t$–$J$ model~\cite{Zaanen.Wu.2008}, as well as its phase-string representation~\cite{Weng_2011,Weng1997,Weng1999}. Question (a) is answered in Sec.~\ref{VMC}. We explain why the low-energy spectral weight undergoes a shift from the nodal to the antinodal region upon changing the sign of $t^\prime$ from negative to positive. Numerical calculations are presented to elucidate the effects of $t^\prime$ on the coherent quasiparticle component and the incoherent component, thereby motivating the introduction of a two-fluid description for $t^\prime>0$. Section~\ref{PhenoGreenFunc} applies this two-fluid picture to the finite-doping regime to establish a direct connection with experimental observations, thereby answering question (b). We start by reviewing the random phase approximation (RPA) analysis of Ref.~\cite{Crossover_Fermi_arc} and then discuss the corresponding phenomenological Green’s function in the electron-doped regime. Finally, in Sec.~\ref{Discussion}, we highlight the distinctive features of our momentum-selective two-fluid picture by comparing it with various existing theoretical approaches.

\section{Revisit: Theoretical Formulation}
\label{RevisitPS}
\subsection{Phase-String Representation of the $t$-$J$ Model}
\label{t-J fractionalization}
In this section, we introduce the theoretical basis that will be used in the following discussion. We begin by introducing the phase-string representation of the $t$-$J$ model on a two-dimensional square lattice. The Hamiltonian is $H_{t\text{-}J}=H_t+H_J$, where
\begin{equation}
\begin{split}
    H_t&=-t\sum_{\braket{ij}\sigma}P_sc^\dagger_{i\sigma}c_{j\sigma}P_s+\text{H.c.},\\
    H_J&=J\sum_{\braket{ij}}P_s(\boldsymbol{S}_i\cdot \boldsymbol{S}_j-\frac{1}{4}n_i n_j)P_s.
\end{split}
\label{HtJeqn}
\end{equation}
Here, $\boldsymbol{S}_i=\frac{1}{2}\sum_{\alpha\beta} c^\dagger_{i\alpha}\boldsymbol{\rho}_{\alpha\beta}c_{i\beta}$ is the local \text{SU(2)} spin operator at site $i$, $\boldsymbol{\rho}=(\rho_x,\rho_y,\rho_z)$ represents three Pauli matrices.  $n_i=\sum_\sigma c^\dagger_{i\sigma}c_{i\sigma}$ is the particle number operator at site $i$, and $P_s$ is the projection operator that enforces the no-double-occupancy constraint.

At half-filling, the charge degrees of freedom are frozen, and the ground state is governed solely by $H_J$, which has been well understood as an AFM long-range-ordered state. Upon hole doping, two central issues remain under intense debate: the propagation dynamics of a single hole within the quantum spin background, and the microscopic mechanism driving hole pairing in the absence of an external pairing force.

For the first issue, early studies treated the doped hole as a coherent Landau quasiparticle whose effective mass is renormalized by holon-magnon scattering, as in the self-consistent Born approximation (SCBA) approach~\cite{SchmittRink1988,PhysRevB.39.6880,MartinezHorsch1991}. This approach reproduces a dispersion relation similar to unbiased numerical results~\cite{PhysRevB.62.15480}. However, recent DMRG studies \cite{PhysRevB.98.165102} on systems with open boundary conditionss and $C_4$ rotational symmetry have revealed hidden spin currents and $2\times 2$ charge loop currents in the single-hole ground state (characterized by the quantum number $L_z = \pm 1$). These findings suggest that spin-charge entanglement and their local motions are far more complex than what is captured by the SCBA. The latter mainly focuses on the long-wavelength regime and neglects local singular quantum-interference effects in the hole's motion. To handle this local effect properly, Ref.~\cite{Zaanen.Wu.2008} analyzed the phase-string sign structure of the $t$-$J$ model: the $H_t$ term generates severe $\pm 1$ phase frustration depending on whether spin-up or spin-down particles are exchanged with the hole during its motion. This nonperturbative frustration can be ``softened'' by introducing a composite fermion (twisted quasiparticle)~\cite{Chen_single_hole},
\begin{equation}
\tilde{c}_{i\sigma} =c_{i\sigma}e^{-i\hat{\Omega}_i},
\end{equation}
where the non-local phase-shift operator $\hat{\Omega}_i$ is defined as
\begin{equation}
    \hat{\Omega}_i = \sum_{l(\neq i)}\theta_i(l)n_{l\downarrow},
\label{Omegaold}
\end{equation}
where $n_{l\downarrow}$ counts the number of spin-$\downarrow$ particles at site $l$, $\theta_i(l)=\pm \mathrm{Im}\ln(z_i-z_l)$, and $z_i$ is the complex coordinate of site $i$. By explicitly twisting the background spins, the twisted hole $\tilde{c}_{i\sigma}$ transforms into a non-Landau quasiparticle that propagates with much greater coherence than the bare hole $c_{i\sigma}$, as demonstrated by VMC studies on two-leg ladders~\cite{Zhao_two_leg} and 2D lattices~\cite{Chen_single_hole}. Crucially, the variational wave function containing the twisted hole $\tilde{c}$ successfully recovers the local spin and charge current patterns~\cite{cancui}, indicating its capability to capture the mutual entanglement between the hole and its spin environment.

For the second issue, DMRG studies indicate that holes bind tightly in real space within a few lattice constants, much shorter than the AFM correlation length \cite{PhysRevB.55.6504}. This stands in contrast to conventional Bardeen–Cooper–Schrieffer (BCS) theory, where pairing arises from a Fermi surface instability mediated by a pairing ``glue" such as AFM fluctuations \cite{PhysRevLett.69.961}. Similar hole-hole correlations and the $d$-wave symmetry of Cooper pairing are reproduced by VMC simulations~\cite{Chen_Two_hole,Zhao_Two_hole}, in which two twisted holes $\tilde{c}$ are found to tightly pair to eliminate the spin current surrounding the individual holes.

These numerical results dictate the introduction of a unitary transformation $e^{i\hat{\Theta}}$~\cite{Weng1997,Weng1999,Ma_2014,Weng_2011} to explicitly account for the singular phase-string sign structure, thereby yielding a ground-state representation that departs fundamentally from conventional slave-particle frameworks~\cite{Kotliar_Liu,SU2gauge,Patrick_Nagaosa,DopeMottInsulator}. The ground state can be expressed as~\cite{Weng_2011,Ma_2014}:
\begin{equation}
\ket{\Psi_G}\equiv e^{i\hat{\Theta}}\ket{\Phi_G}    
\end{equation}
\begin{equation}
    \ket{\Phi_G}=\hat{P}\ket{\Phi_{\tilde{c}}}\otimes\ket{\Phi_b},
\label{GS in phase string}
\end{equation}
where $\ket{\Phi_{\tilde c}}$ corresponds to a fermionic $s$-wave (nodeless) BCS-like paired state of the  composite fermion $\tilde{c}$, while $\ket{\Phi_b}$ describes a short-range resonating valence bond (RVB) state.

The unitary transformation is defined as
\begin{equation}
    \hat{\Theta}=-\sum_i n^{\tilde{c}}_i \hat{\Omega}_i,
    \label{duality transformation}
\end{equation}
where $n_i^{\tilde{c}}$ denotes the $\tilde{c}$ number at site $i$. Equation~\eqref{duality transformation} admits a physical interpretation in which each doped hole at site $i$ introduces a nonlocal twist $\hat{\Omega}_i$ of the $b$-spinon background, with
\begin{equation}
    \hat{\Omega}_i = \frac{1}{2}(\Phi_i^s-\Phi_i^0),
\label{Omega}
\end{equation}
\begin{equation}
    \Phi_i^s=\sum_{l(\neq i)}\theta_i(l)(n_{l\uparrow}^b-n_{l\downarrow}^b),
\end{equation}
\begin{equation}
    \Phi_i^0=\sum_{l(\neq i)}\theta_i(l).
\end{equation}

The definition of $\hat{\Omega}_i$ in Eq.~\eqref{Omega} naturally extends Eq.~\eqref{Omegaold} to the present ground-state representation, under the constraint imposed by the projection operator $\hat{P}$, which enforces two local constraints:
\begin{equation}
    \sum_{\sigma}n^b_{i\sigma}=1,\quad n^{\tilde{c}}_{i\bar\sigma}=n^{\tilde{c}}_in^{b}_{i\sigma}.
\end{equation}
The first condition ensures that, in $\ket{\Phi_b}$, the bosonic $b$-spinon sector is constrained to be half-filled at every site. The second condition imposes a local singlet constraint between the fermionic $\tilde c$ in $\ket{\Phi_{\tilde{c}}}$ and the $b$-spinons. Namely, a $\tilde c$ fermion with spin $\bar\sigma=-\sigma$ is bound to 
a $b$-spinon with spin $\sigma$, thereby neutralizing the original local moment at a hole site, where $n^{\tilde{c}}_i=1$. In this sense, doping a half-filled Mott insulator is equivalent to introducing additional itinerant spin degrees of freedom that quench the original local moments.

A mean-field calculation can be carried out based on the fractionalization scheme in Eq.~\eqref{GS in phase string}. The corresponding effective Hamiltonians governing the dynamics of these fractionalized degrees of freedom are briefly discussed in Appendix~\ref{Effective Hamiltonian}.

 
\subsection{Interference Frustration Due to Next-Nearest-Neighbor Hopping $t^\prime$}
\label{Sec of sign structure}
The NNN hopping term,
\begin{equation}
H_{t'}=-t'\sum_{\braket{\braket{ij}},\sigma} P_s c^\dagger_{i\sigma} c_{j\sigma} P_s + \text{H.c.},
\label{HttpJ}
\end{equation}
is often introduced to break the particle-hole symmetry and distinguish hole doping from electron doping through the sign of $t'$. The total Hamiltonian then reads $H_{t\text{-}t'\text{-}J} = H_t + H_{t'} + H_J$, where $H_t$ and $H_J$ are defined in Eq.~\eqref{HtJeqn}.

On a square lattice, the hopping Hamiltonian $H_t + H_{t'}$ remains invariant under the sign change $t \rightarrow -t$ when combined with the sublattice transformation $c^\dagger_{i\sigma} \rightarrow (-1)^i c^\dagger_{i\sigma}$. On the other hand, a sign flip of $t'$ alone ($t' \rightarrow -t'$) can be physically mapped to the particle-hole transformation $c^\dagger_{i\sigma} \rightarrow (-1)^i c_{i\sigma}$. This transformation maps a hole-doped system with $t' < 0$ directly onto an electron-doped system with $t' > 0$ at the corresponding doping level. Consequently, one can systematically explore the electron-doped regime by simply reversing the sign of $t'$ while staying in the hole-doped language. This mapping is discussed in further detail in Appendix~\ref{sign of t'}.

In the following, we show that the sign of $t^\prime$ can generate distinct quantum-interference effects for doped charges in the presence of a spin background in doped Mott insulators. The generic partition function $Z=\operatorname{Tr} e^{-\beta H}$ can be expanded as a sum over closed paths of imaginary-time evolution,
\begin{equation}
Z=\sum_{n=0}^{\infty} \sum_{\left\{\alpha\right\}_n} \frac{\beta^{n}}{n !} \prod_{k=0}^{n-1} \left\langle\alpha_{k+1}\right| (-H) \left| \alpha_{k}\right\rangle,
\label{partition function expansion}
\end{equation}
where each $\ket{\alpha_k}$ belongs to a complete basis of the Hilbert space, such as the real-space Fock basis. Each sequence of states $\{\alpha\}_n$ satisfies the temporal periodic boundary condition $\ket{\alpha_{n}}=\ket{\alpha_{0}}$. Non-zero matrix elements $\left\langle\alpha_{k+1}|(-H)| \alpha_{k}\right\rangle$ describe single steps in the physical evolution, and each loop consisting of $n$ matrix elements represents a closed $n$-step worldline. Importantly, each evolution step in a quantum many-body system can carry a phase (sign) rather than being purely positive. After completing an arbitrary worldline loop, the accumulated phase (sign) is gauge invariant. 

A rigorous and systematic analysis of the $t$-$t'$-$J$ model has been presented in Refs.~\cite{Zaanen.Wu.2008, PhysRevB.110.165127} and revisited in Appendix~\ref{sign structure}. Here, to minimally illustrate the effect of the sign of $t^\prime$ on the propagation of a doped hole, we analyze the contribution to the partition function from a minimal worldline, namely a triangular hopping loop on a square lattice with an AFM spin background, as illustrated in Fig.~\ref{sign structure triangular}.
The local Hilbert space with the no-double-occupancy constraint is $\{| \circ\rangle, |\uparrow \rangle, |\downarrow \rangle\}$, and each evolution step appearing in Eq.~\eqref{partition function expansion} can be represented in this basis by the following matrix elements:

\begin{figure}[t]
	\centering
	\includegraphics[width=0.95\linewidth]{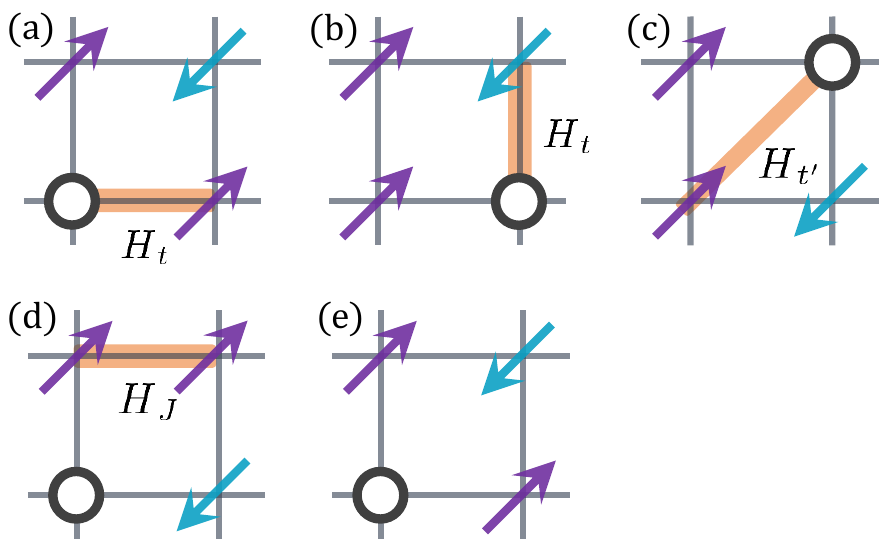}
	\caption{Panels (a)--(e) illustrate a hopping process around the minimal triangular loop on an antiferromagnetic background. In the first step, a hole exchanges positions with an up spin and subsequently with a down spin via the nearest-neighbor hopping channel $t$. It then exchanges with another spin through the next-nearest-neighbor hopping channel $t^\prime$. Finally, the two remaining adjacent spins exchange positions, restoring the plaquette to its original configuration. Under the Ising basis, the entire process from (a) to (e) contributes a sign proportional to $\mathrm{sgn}(t^\prime)$.}
	\label{sign structure triangular}
\end{figure}
\begin{equation}
\begin{split}
    \bra{\circ\uparrow}(-H_t)\ket{\uparrow\circ}&=t,\\
    \bra{\circ\downarrow}(-H_t)\ket{\downarrow\circ}&=t,\\
    \bra{\circ\uparrow}(-H_{t^\prime})\ket{\uparrow\circ}&=-t^\prime,\\\bra{\circ\downarrow}(-H_{t^\prime})\ket{\downarrow\circ}&=-t^\prime,\\
    \bra{\uparrow\downarrow}(-H_J)\ket{\downarrow\uparrow}&=-\frac{J}{2}.\\
\end{split}
\end{equation}
Here the NN density-density interaction is neglected in the calculation since it is diagonal in our chosen basis and is unrelated to the spin fluctuation. Thus, the contribution of such a loop to the partition function is $t\times t\times(-t^\prime)\times (-J/2)$, with its sign determined by $\mathrm{sgn}(t^\prime)$.
As a consequence, constructive interference among worldlines is more strongly favored when $t^\prime>0$, allowing hole motion with a reduced tendency to disrupt the AFM background. This qualitative argument provides a possible explanation for the experimentally observed enhancement of AFM order in electron-doped cuprates, which is similar to the mechanism of the counter-Nagaoka effect~\cite{Glittum2025,Sriram2005}. Moreover, this strongly suggests that the doped carrier propagates much more coherently on the electron-doped side ($t'>0$) than on the hole-doped side ($t'<0$), which provides the central physical motivation for the following analysis.



\section{VMC Study of $t^\prime$-Induced Asymmetry}
\label{VMC}
To elucidate how the sign of the NNN hopping parameter $t'$ affects the low-energy carrier properties, we perform VMC calculations for the single- and two-hole-doped $t$-$t'$-$J$ model given in Eq.~\eqref{HttpJ}. We mainly focus on the single-hole case and briefly discuss the two-hole case at the end of this section. Following the formalism in Eq.~\eqref{GS in phase string}, we start from the half-filled ground state $\ket{\phi_0}$ with AFLRO, which is well captured by the Liang-Doucot-Anderson-type variational wave function \cite{Liang1988}. Rather than introducing the bare doped hole $c_i$, we introduce a composite fermion $\tilde{c}_i \equiv c_i e^{\mp i\hat{\Omega}_i}$ into this background. The single-hole-doped~\footnote{In the electron-doped case, a doped electron is mapped onto a ``hole'' via a particle-hole transformation. Thus, the $t$-$t'$-$J$ model remains inherently hole-doped in our description, where $t'>0$ and $t'<0$ are used to characterize the electron-doped and hole-doped regimes, respectively.} wave-function \emph{Ansatz} is constructed as follows:
\begin{equation}
|\Psi_G\rangle_{1\mathrm{h}}=\sum_{i,v,m=\pm 1}\varphi_{m}(i,v)c_{i\uparrow}e^{-im\left(\hat{\Omega}_{i}-\hat{\Omega}_{v}\right)}|\phi_{0}\rangle, \label{singlehole}
\end{equation}
where the $\varphi_{m}(i,v)$ are variational parameters determined using the VMC method, and $e^{\pm i \hat{\Omega}_{v}}$ represents an antivortex excitation at a plaquette center $v$, which compensates the logarithmic divergence in the superexchange energy caused by the spin-current vortex surrounding the composite fermion $\tilde{c}_{i}$ \cite{Chen_single_hole}.

In our VMC simulations, we fix $t/J = 2$ and employ open boundary conditions. Benchmark density matrix renormalization group (DMRG) calculations confirm the VMC results, showing close agreement in both ground-state energies and quasiparticle spectral weights (see Appendix~\ref{numerical_results}).

\subsection{Evolution of the Ground-State Energy}
Importantly, Eqs.~\eqref{GS in phase string},  and \eqref{singlehole}, imply that the twisted hole $\tilde{c}_{i\sigma}$ behaves as a nearly plane-wave-like object,
while the physical quasiparticle $c_{i\sigma}=\tilde{c}_{i\sigma}e^{\pm i\hat{\Omega}_i}$ arises from the recombination process between the twisted hole $\tilde{c}_{i\sigma}$ and the antivortex $e^{\pm i \hat{\Omega}_{v}}$, or equivalently, when the hole--antivortex distance $|i-v|$ is small. Therefore, from the perspective of the physical hole probed in spectroscopic measurements, this structure implies that a substantial incoherent component is inevitably present, even in the presence of a potentially coherent quasiparticle contribution. Accordingly, the single-hole ground state $|\Psi_G\rangle_{1\mathrm{h}}$ in Eq.~\eqref{singlehole} can be naturally decomposed into two orthogonal components:
\begin{equation}
    |\Psi_G\rangle_{1\mathrm{h}} = \eta_1|\Psi_{\text{qp}}\rangle_{1\mathrm{h}} + \eta_2|\Psi_{\text{ic}}\rangle_{1\mathrm{h}}.
    \label{1h_component}
\end{equation}

Here, $|\Psi_{\text{qp}}\rangle_{1\mathrm{h}} =\sum_i\varphi(i)c_{i\uparrow} |\phi_0\rangle$ only contains the bare quasiparticle, while the many-body twist effect from the $e^{\mp i\hat{\Omega}_i}$ is captured by the incoherent component $|\Psi_{\text{ic}}\rangle_{1\mathrm{h}}$. The orthogonality between the two components $_{1\mathrm{h}}\langle \Psi_{\text{qp}} | \Psi_{\text{ic}} \rangle_{1\mathrm{h}} = 0$ is achieved by imposing $_{1\mathrm{h}}\langle \Psi_{\text{ic}} | c_{i\uparrow}| \phi_0 \rangle = 0$ for all sites $i$. Under this condition, $\eta_1 \varphi(i) = 2\langle \phi_0 | c_{i\uparrow}^{\dagger} |\Psi_G\rangle_{1\mathrm{h}}$ can be determined. With $|\Psi_{\text{qp}}\rangle_{1\mathrm{h}}$ and $|\Psi_{\text{ic}}\rangle_{1\mathrm{h}}$ normalized to unity, the coefficient satisfies $|\eta_1|^2+|\eta_2|^2 = 1$. The total kinetic energy is then decomposed as
\begin{equation}
    E^{1\mathrm{h}}_{K} = {}_{1\mathrm{h}}\!\langle \Psi_G | \hat H_{K} | \Psi_G \rangle_{1\mathrm{h}}=E^{\text{qp}}_{K}+E^{\text{ic}}_{K}+E^{\text{cr}}_{K} \label{1hEt}
\end{equation}
where $H_{K} = H_t + H_{t'}$ is the total kinetic Hamiltonian, $E^{\text{qp}}_{K}$, $E^{\text{ic}}_{K}$ and $E^{\text{cr}}_{K}$ denote the kinetic energy gain from the propagation of $c_{i}$, $\tilde{c}_{i}$, and the resonance between them, respectively, as defined below:
\begin{eqnarray}
    \begin{aligned}
    E^{\text{qp}}_{K} &\equiv |\eta_1|^2  {}_{1\mathrm{h}}\left\langle\Psi_{\mathrm{qp}}\right| \hat{H}_K\left|\Psi_{\mathrm{qp}}\right\rangle_{1\mathrm{h}}\\
    E^{\text{ic}}_{K} &\equiv |\eta_2|^2 {}_{1\mathrm{h}} \langle \Psi_{\text{ic}}|\hat{H}_K|\Psi_{\text{ic}}\rangle_{1\mathrm{h}}\\
    E^{\text{cr}}_{K} &\equiv \eta_1^*\eta_2 {}_{1\mathrm{h}} \langle \Psi_{\text{qp}}|\hat{H}_K|\Psi_{\text{ic}}\rangle_{1\mathrm{h}} +\mathrm{H.c.}
    \end{aligned}
    \label{1h_3comp_Et}
\end{eqnarray}

For a momentum-space representation $|\Psi_{\text{qp}}\rangle_{1\mathrm{h}}=\sum_{\boldsymbol{k}}\varphi(\boldsymbol{k})c_{\boldsymbol{k}\uparrow} |\phi_0\rangle$, which contains the bare quasiparticle $c_{\boldsymbol{k}}$ with amplitude $\varphi(\boldsymbol{k})$, the quasiparticle kinetic energy is given by $ E^{\text{qp}}_{K}= |\eta_1|^2\sum_{\boldsymbol{k}} |\varphi(\boldsymbol{k})|^2 \epsilon^c_{\boldsymbol{k}}$, where the quasiparticle dispersion reads
\begin{equation}
    \epsilon^{c,\mathrm{eff}}_{\boldsymbol{k}} = -2t_{\mathrm{eff}}(\cos k_x+\cos k_y)+4t_{\mathrm{eff}}^{\prime}\cos k_x\cos k_y .
    \label{Et1hqp}
\end{equation}
Here, $t_{\mathrm{eff}}=-t
\bra{\phi_0}(n_{i\uparrow}n_{j\uparrow}+S^+_iS^-_j)\ket{\phi_0}$ ($i=\text{NN}(j)$) and $t^{\prime}_{\mathrm{eff}}=t^{\prime}\bra{\phi_0}(n_{i\uparrow}n_{j\uparrow}+S^+_iS^-_j)\ket{\phi_0}$ ($i=\text{NNN}(j)$) denote the renormalized hopping parameters, which are found to retain the same signs as the bare $t$ and $t^{\prime}$, respectively.

\begin{figure}[t]
\centering
\includegraphics[width=0.8\linewidth]{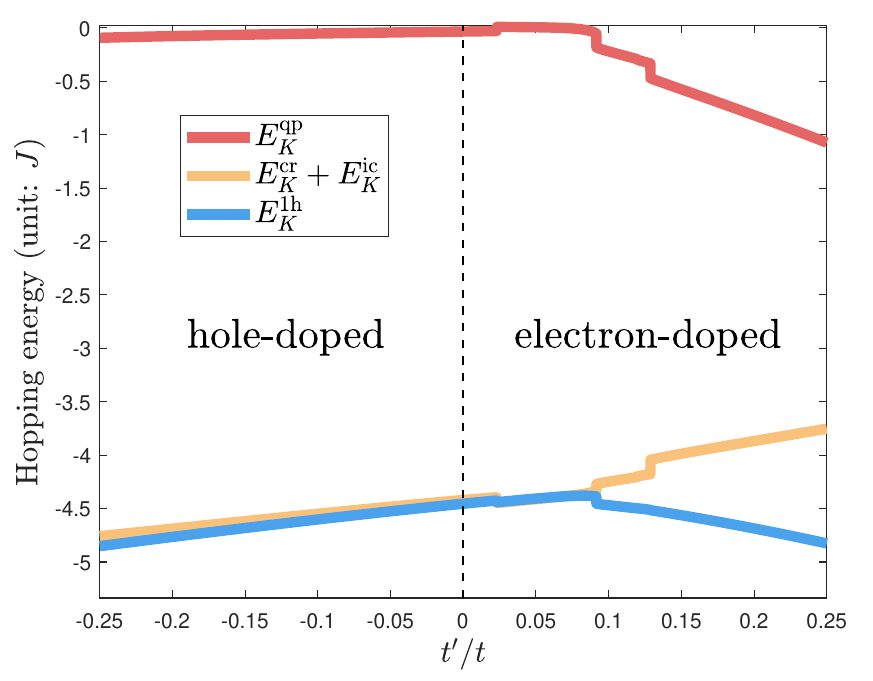}
\caption{Kinetic energy components $E^{\text{qp}}_{K}$, $E^{\text{ic}}_{K}+E^{\text{cr}}_{K}$ and $E^{1\mathrm{h}}_{K}$ (defined in Eq.~\eqref{1hEt}) as functions of $t'/t$, computed using VMC on a $12\times 12$ lattice. The abrupt jumps originate from level crossings.}
\label{Etcomponents}
\end{figure}

To elucidate how different kinetic-energy contributions evolve with $t'/t$, Fig.~\ref{Etcomponents} shows the VMC results for $E^{\text{qp}}_{K}$, $E^{\text{ic}}_{K}+E^{\text{cr}}_{K}$, and the total $E^{1\mathrm{h}}_{K}$ as a function of $t'/t$. Changes in the superexchange energy with $t'/t$ are much smaller than those in the kinetic energy and are therefore not considered here.
For $t'/t \leq 0$, corresponding to the hole-doped regime, the quasiparticle kinetic energy $E^{\text{qp}}_{K}$ is nearly zero and contributes negligibly to $E^{1\mathrm{h}}_{K}$. Consequently, the kinetic energy is almost entirely provided by the motion of $\tilde{c}_{i}$ and its resonance with $c_{i}$: $E^{\text{ic}}_{K}+E^{\text{cr}}_{K} \approx E^{1\mathrm{h}}_{K}$. This indicates that the quasiparticle here does not propagate coherently as a fundamental object, but rather gains its mobility and kinetic energy via a resonance with the incoherent sector. Physically, this implies that, in the hole-doped case, the quasiparticle emerges as a bound state formed by the composite fermion $\tilde{c}_i$ and an antivortex excitation $e^{\pm i \hat{\Omega}_{v}}$, rather than being an intrinsic ``building block'' as in a conventional Fermi liquid.

In contrast, for $t'/t > 0$ (electron-doped regime), the quasiparticle kinetic energy $E^{\text{qp}}_{K}$ decreases significantly beyond a transition at $t'/t = 0.09$, eventually reaching a substantial fraction of $E^{1\mathrm{h}}_{K}$. Although $E^{\text{ic}}_{K}+E^{\text{cr}}_{K}$ increases with $t'/t$, this rise is offset by the growing kinetic energy gain from the quasiparticle motion, leading to a net reduction in $E^{1\mathrm{h}}_{K}$. This indicates that, in the electron-doped regime, beyond a threshold value of $t^\prime/t$, the quasiparticle can acquire kinetic energy via two separate mechanisms, namely a resonance with the incoherent sector, as in the hole-doped case, and an additional intrinsic propagation channel, which endows the quasiparticle with its own coherent dynamical behavior.

\begin{figure*}[t]
\centering
\includegraphics[width=0.90\linewidth]{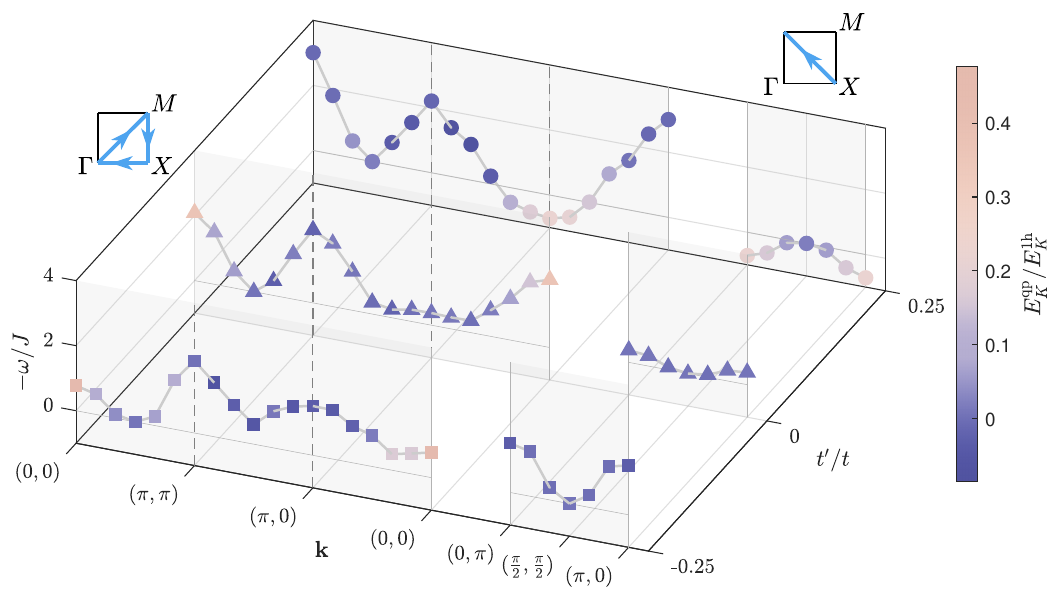}
\caption{Quasiparticle dispersion $\omega(\boldsymbol{k})$ and the ratio 
$\mathcal{C}(\boldsymbol{k})=E^{\mathrm{qp}}_{K}(\boldsymbol{k})/E^{1\mathrm{h}}_{K}(\boldsymbol{k})$ obtained from VMC simulations on a 
$12\times 12$ lattice with $t' = 0,\pm 0.25t$. For each momentum point $\boldsymbol{k}$, the dispersion $\omega(\boldsymbol{k})$ is extracted from the maximum of the spectral function $A^n(\boldsymbol{k},\omega)$, and the corresponding state with the largest spectral weight is used to evaluate the kinetic-energy ratio $\mathcal{C}(\boldsymbol{k})$. To make the vertical axis follow the usual convention of band dispersion, we use $-\omega/J$ on the ordinate.   }
\label{EkVMC}
\end{figure*}

\subsection{Momentum-Dependent Quasiparticle Dynamics}
\label{Energy band asymmetry induced by t'}
To further investigate the asymmetry in $E^{\mathrm{qp}}_{K}$ in momentum space, we calculate the single-particle spectral function defined below:
\begin{equation}
A^-(\boldsymbol{k},\omega)=-\mathrm{Im}\sum_{n}\frac{\left|\langle\Psi_{\mathrm{1h}}(n)|c_{\boldsymbol{k}\uparrow}|\phi_{0}\rangle\right|^{2}}{\omega+[E_{\mathrm{1h}}(n)-E_{\mathrm{0h}}^G+\mu_{-}]+i\eta},
\label{negbias}
\end{equation}
where $|\Psi_{\mathrm{1h}}(n)\rangle$ is the $n$-th excited state in the single-hole sector with eigenenergies $E_{\mathrm{1h}}(n)$ obtained from the ansatz in Eq.~\eqref{singlehole}. $E_{\mathrm{0h}}^G$ is the half-filled ground-state energy, $\mu_{-}$ is the chemical potential, and $\eta = 0.05J$ is the broadening parameter. The quasiparticle dispersion $\omega(\boldsymbol{k})$ is defined as the energy at which the spectral function $A^-(\boldsymbol{k},\omega)$ is maximized. For each momentum point, we identify among all eigenstates the one whose energy lies closest to $\omega(\boldsymbol{k})$ and carries the largest spectral weight $\left|\langle\Psi_{\mathrm{1h}}(n)|c_{\boldsymbol{k}\uparrow}|\phi_{0}\rangle\right|^{2}$. For this state, we can decompose it into quasiparticle and incoherent components (cf. Eq.~\eqref{1h_component}), calculate the corresponding bare-quasiparticle kinetic energy $E^{\mathrm{qp}}_{K}(\boldsymbol{k})$ (cf. Eq.~\eqref{1h_3comp_Et}), and evaluate the ratio $\mathcal{C}(\boldsymbol{k})=E^{\mathrm{qp}}_{K}(\boldsymbol{k})/E^{1\mathrm{h}}_{K}(\boldsymbol{k})$, which quantifies the relative contribution of the bare quasiparticle hopping process to the total kinetic energy. By analyzing $\mathcal{C}(\boldsymbol{k})$, we can distinguish whether a large quasiparticle weight of $c_{\boldsymbol{k}}$ on the energy band originates from intrinsic quasiparticle propagation or from the resonance-induced process. The momentum-resolved dispersion $\omega(\boldsymbol{k})$, together with the ratio $\mathcal{C}(\boldsymbol{k})$, therefore provides a direct measure of how the enhancement of the quasiparticle's intrinsic motion becomes momentum-selective for $t'>0$. The resulting dispersions $\omega(\boldsymbol{k})$ and kinetic-energy ratios $\mathcal{C}(\boldsymbol{k})$ for $t'=-0.25t, 0$, and $0.25t$ are shown in Fig.~\ref{EkVMC}.


In Fig.~\ref{EkVMC}, since the ground-state energies differ slightly for different $t'$ values, we adjust the chemical potential $\mu_-$ separately for each $t'/t$ value such that $\omega(\boldsymbol{k} = (\pi/2,\pi/2)) = 0$. This choice is made solely for clarity: at $t' =0$, the band minima are located at $(\pm\pi/2,\pm\pi/2)$, and this reference point allows a direct comparison of how finite $t'$ lifts or lowers the dispersion in the antinodal region and near the point $(0,0)$ without altering the band dispersion structure. We first note that at $t' = 0$, the dispersion pattern closely matches unbiased Green-function Monte Carlo results \cite{BONINSEGNI1994330}, validating the ansatz in Eq.~\eqref{singlehole}. In this case, the energies in the antinodal region are higher than those near the nodal points by an energy difference $\Delta E_{d}\sim 0.38J$.

For $t'= -0.25t$ (hole-doped regime), the band minimum remains located at $(\pm\pi/2,\pm\pi/2)$ while the contribution from bare quasiparticle hopping is nearly zero, with $\mathcal{C}(\boldsymbol{k}) \approx 0$.
This indicates that the low-lying quasiparticle mode in the hole-doped case resides in the nodal region, and its propagation is dominated by the resonance-induced process. The intrinsic propagation of the quasiparticle itself is negligible, and the low-energy physics is therefore governed by the incoherent sector.

By contrast, for $t'= 0.25t$ (electron-doped regime), the band minima shift to the points $(\pi,0)$ and $(0,\pi)$, and low-energy excitations in the antinodal regions exhibit substantial kinetic-energy contributions from quasiparticle motion, as evidenced by the large ratio $E^{\mathrm{qp}}_{K}/E^{1\mathrm{h}}_{K} \sim 0.22$. This behavior aligns with the dispersion relation $\epsilon^{c,\mathrm{eff}}_{\boldsymbol{k}}$ in Eq.~\eqref{Et1hqp}, where the $t_{\mathrm{eff}}^{\prime}$ term reaches its minimum in the antinodal region for $t_{\mathrm{eff}}^{\prime}>0$. Thus, NNN hopping of the bare quasiparticle drives the significant reduction in $E^{\mathrm{qp}}_{K}$ with increasing $t'/t$ (shown in Fig.~\ref{Etcomponents}), overcoming the energy cost $\approx \Delta E_{d}$ and driving the shift of the band minima from $(\pm\pi/2,\pm\pi/2)$ to $(\pi,0)$ and $(0,\pi)$. Moreover,  the quasiparticle modes near the nodal region also exhibit a modification of the dispersion, with their energies shifted to higher values as $t'$ increases. However, the ratio $\mathcal{C}(\boldsymbol{k})$ in the nodal region changes only weakly, indicating that the contribution from bare quasiparticle hopping remains small and does not become comparable to that in the antinodal region.
These behaviors imply that, in the electron-doped case at the single-hole limit, the strongly correlated character of the low-energy modes in the antinodal region is substantially reduced and is replaced by a more conventional quasiparticle behavior, which can be effectively captured by a band-like description. By contrast, the nodal region continues to be dominated by resonance-induced dynamics. Although the corresponding mode resides at relatively high energy in the single-hole limit, it can evolve into a low-lying excitation at finite doping, as will be demonstrated by the following phenomenological theory. The distinct behaviors between the antinodal and nodal regions therefore suggest qualitatively different quasiparticle dynamics in these two momentum-separated regimes.


\subsection{Antinodal Quasiparticle as an Independent Mode}
To further elucidate the distinct quasiparticle nature in the electron-doped and hole-doped cases, we construct a conventional reference for comparison in the following. Specifically, we consider a Bloch-wave ansatz in which the bare quasiparticle operator $c_{i\sigma}$ is doped into the spin background without fractionalization:
\begin{equation}
|\Psi_{\text{Bloch}}\rangle_{1\mathrm{h}}=\sum_{i}\varphi_{\mathrm{B}}(i)c_{i\uparrow}|\phi_{0}\rangle. \label{Bloch}
\end{equation}
Using its variational ground state, we compute the quasiparticle spectral weight
$Z_{\boldsymbol{k}}=|\langle\phi_0|c^{\dagger}_{\boldsymbol{k}}|\Psi_{\text{Bloch}}\rangle_{1\mathrm{h}}|^2$ (see Fig.~\ref{zkVMC}(c)(d) in Appendix~\ref{numerical_results}), and compare the resulting momentum dependence with that obtained from our wave-function ansatz in Eq.~\eqref{singlehole}, which incorporates incoherent sectors, as well as with unbiased DMRG results.

For $t'=-0.25t$, this Bloch-wave ansatz yields a qualitatively incorrect momentum distribution, with the spectral weight erroneously peaked at $(0,0)$. This failure demonstrates that the quasiparticle modes at $(\pm\pi/2,\pm\pi/2)$, which are unambiguously established by DMRG, cannot be understood as arising from the free propagation of a bare quasiparticle. Instead, they originate from a nontrivial interplay between the quasiparticle and the incoherent background, encoded in the resonance between the composite fermion $\tilde{c}_i$ and the antivortex excitation $e^{\pm i \hat{\Omega}_{v}}$, which gives rise to the nontrivial momentum shift.

For $t' = 0.25t$, the calculated $Z_{\boldsymbol{k}}$ exhibits pronounced peaks at $(\pi,0)$ and $(0,\pi)$, in agreement with the results obtained from $|\Psi_G\rangle_{1\mathrm{h}}$ in Eq.~\eqref{singlehole} as well as from DMRG. The fact that the quasiparticle momenta are correctly captured by the unfractionalized Bloch-wave ansatz further indicates that the low-lying quasiparticles in the electron-doped case, particularly in the antinodal region, exhibit a more conventional character. In this regime, the additional incoherent component in $|\Psi_G\rangle_{1\mathrm{h}}$ primarily serves to further optimize the electron kinetic energy while leaving the quasiparticle momenta unchanged. This role is analogous to the dressing of a Landau quasiparticle by an electron--hole cloud.

Taken together, an important conclusion can be drawn from the single-hole–doped $t$–$t^\prime$–$J$ model. For $t^\prime<0$, the kinetic-energy gain is dominated by the resonance between the quasiparticle and incoherent modes, giving rise to low-lying excitations near the nodal region, while the direct contribution from bare quasiparticle propagation is negligible. By contrast, for $t^\prime>0$, the kinetic-energy contribution associated with the quasiparticle mode is strongly enhanced. This indicates that the intrinsic propagation of the bare quasiparticle emerges as an additional, independent effective channel in determining the low-energy quasiparticle spectrum, particularly in the antinodal region.


Synthesizing the insights from the foregoing analysis with the $t'=0$ ground-state wave function [Eq.~\eqref{GS in phase string}], a unified picture emerges for the ground state and low-energy excited states across different signs of $t^\prime$. This comprehensive description, which encapsulates the stark contrast between the electron- and hole-doped regimes, is outlined below:
\begin{subequations}
\begin{align}
\ket{\Psi_{t^\prime<0}}
&= \ket{\Psi_G},\\
\ket{\Psi_{t^\prime>0}}
&= \ket{\Psi_c}\otimes \ket{\Psi_G}.
\end{align}
\label{fractionalization in electron}
\end{subequations}
Here, the wave-function structure in the hole-doped regime ($t'<0$) remains identical to that in the $t'=0$ limit. By contrast, an additional component, $\ket{\Psi_c}$, emerges in the electron-doped regime ($t'>0$). This extra contribution stems from the bare electron propagation becoming a coherent and independent channel, which allows the quasiparticle itself to act as an additional degree of freedom governed by its own intrinsic dynamics.

\subsection{The pairing structure with different signs of $t^{\prime}$}
\label{Two-hole VMC}

Finally, we consider the pairing structure of two holes with different signs of the NNN hopping parameter $t^{\prime}$. The two-hole wave-function \emph{Ansatz} is constructed as follows, where a second hole is added at the antivortex position of the single-hole wave function in Eq.~\eqref{singlehole}, and the position $v$ is now shifted to a lattice site $j$:
\begin{equation}
|\Psi_{\mathrm{G}}\rangle_{2\mathrm{h}}=\sum_{i,j,m=\pm 1}g_{m}(i,j)c_{i\uparrow}c_{j\downarrow}e^{-im\left(\hat{\Omega}_{i}-\hat{\Omega}_{j}\right)}|\phi_{0}\rangle, \label{eq:twohole}
\end{equation} 
Similarly to the single-hole case, we can also decompose the two-hole ground state into two orthogonal components:
\begin{equation}
    |\Psi_G\rangle_{2\mathrm{h}} = \eta_1^{\prime} |\Psi_{\text{qp}}\rangle_{2\mathrm{h}} + \eta_2^{\prime} |\Psi_{\text{ic}}\rangle_{2\mathrm{h}}.
    \label{2h_component}
\end{equation}
where $|\Psi_{\text{qp}}\rangle_{2\mathrm{h}}$ takes the form $\sum_{i,j}\varphi(i,j)c_{i\uparrow}c_{j\downarrow} |\phi_0\rangle$, and the incoherent component $|\Psi_{\text{ic}}\rangle_{2\mathrm{h}}$ is orthogonal to it. The two components are normalized to unity, with the coefficients satisfying $|\eta_1^{\prime}|^2+|\eta_2^{\prime}|^2 = 1$.

\begin{figure}[t]
\centering
\includegraphics[width=0.9\linewidth]{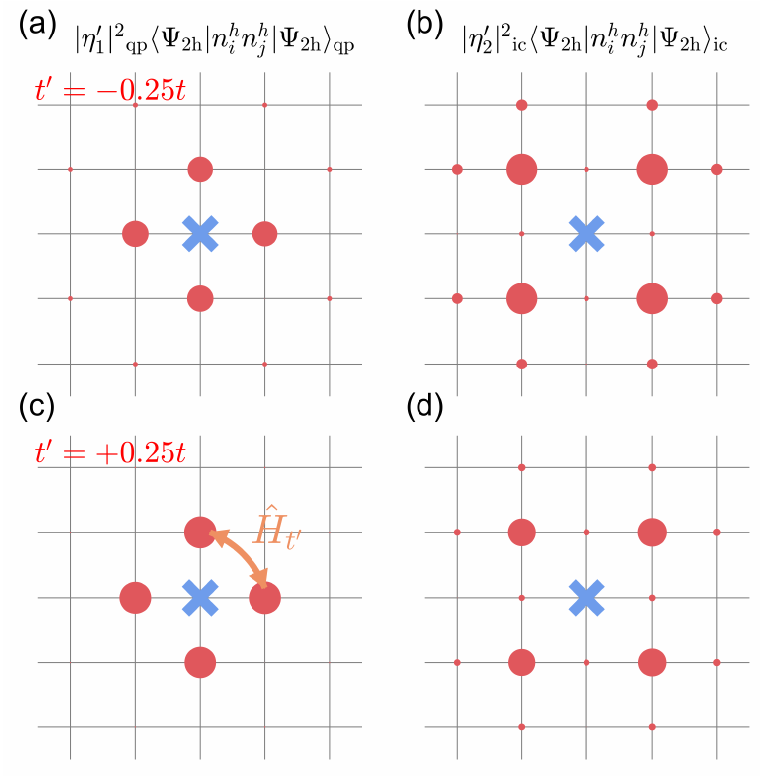}
\caption{Expectation value of the hole-hole correlator $\langle\Psi| n^h_i n^h_j |\Psi\rangle$ evaluated for the two components of the two-hole ground state on a $12\times 12$ system, $|\Psi\rangle = |\Psi_{\text{qp}}\rangle_{2\mathrm{h}}$ and $|\Psi_{\text{ic}}\rangle_{2\mathrm{h}}$. Panels (a) and (b) correspond to $t^{\prime} = -0.25t$, while panels (c) and (d) correspond to $t^{\prime} = 0.25t$. The sizes of the red points represent the relative magnitude of the correlator, where one hole position $i$ is fixed at the blue cross. For comparison between the $t^{\prime} = \pm 0.25t$ cases, the correlator is normalized by $_{2\mathrm{h}}\langle\Psi_G| n^h_i |\Psi_G\rangle_{2\mathrm{h}}$ so that differences in the relative weight of the quasiparticle component can be clearly shown.}
\label{holehole}
\end{figure}

Fig.~\ref{holehole} shows the hole-hole correlation $n^h_i n^h_j$ of the two components for $t' = \pm 0.25t$, where $n_i^h=1-n_{i}$. We find that in both the electron-doped regime ($t^{\prime} = 0.25t$) and the hole-doped regime ($t^{\prime} = -0.25t$), the two holes are tightly paired with a pair size much smaller than the diverging AFM correlation length. Consistent with the $t' = 0$ results in Ref.~\cite{cancui}, their pairing structures show almost no qualitative difference: the quasiparticle component contains the NN Cooper pairing with $d_{x^2-y^2}$ symmetry, and the incoherent component contains the NNN $d_{xy}$ pairing, accompanied by the transverse spin distortion from the many-body twist operator $e^{\mp i(\hat{\Omega}_i -\hat{\Omega}_j ) }$. The tight pairing of the two twisted holes can weaken the distortion to the spin background, while most of the kinetic energy gain from the NN hopping $\hat{H}_t$ can also be achieved by the resonance between the two components.

The Cooper-pair component has a larger weight in the electron-doped regime (with $|\eta_1^{\prime}|^2 = 0.364 $ for $t^{\prime} = -0.25t$ and $|\eta_1^{\prime}|^2 = 0.518$ for $t^{\prime} = 0.25t$). This occurs because the Cooper-pair component can gain kinetic energy through the NNN hopping channel (denoted by the orange arrow in Fig.~\ref{holehole}) when $t'>0$ ($|\eta_1^{\prime}|^2  {}_{2\mathrm{h}}\left\langle\Psi_{\mathrm{qp}}\right| \hat{H}_{t^{\prime}}\left|\Psi_{\mathrm{qp}}\right\rangle_{2\mathrm{h}} = -0.88J$ for $t^{\prime} = 0.25t$ and $= 0.39J$ for $t^{\prime} = -0.25t$). Therefore, once generated from the resonance process, the Cooper pair acquires another hopping channel: it can move much more coherently through the NNN hopping channel, whereas in the hole-doped case, the NNN hopping energy of the Cooper pair component is unfavorable.

\begin{figure*}[ht!]
\centering
\includegraphics[width=0.95\linewidth]{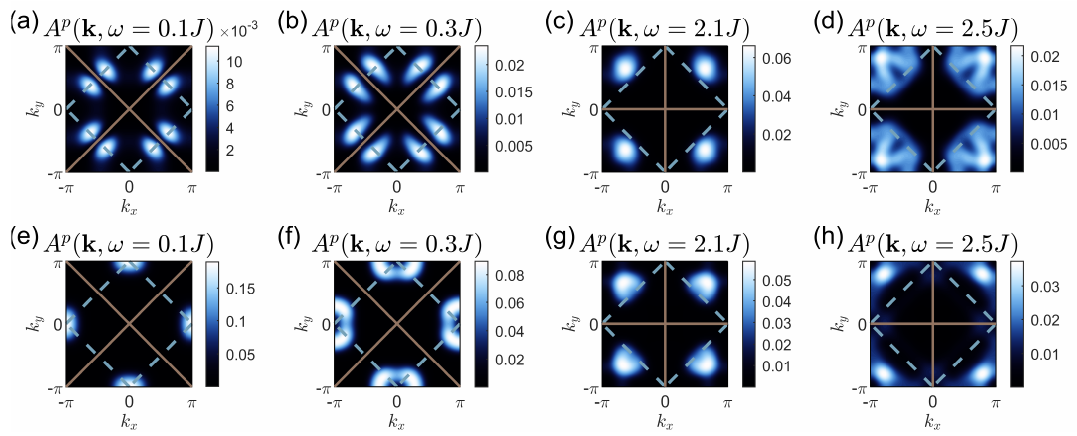}
\caption{Positive bias single-particle spectral function $A^+(\boldsymbol{k},\omega)$ Eq.~\eqref{eq:Ap} for two-hole ground state on a $12\times 12$ system, calculated with $t^{\prime}  = -0.25t$ (panels (a)-(d)) and $t^{\prime}  = 0.25t$ panels (panels (e)-(h)). The brown lines denote the nodal lines of the $d_{x^2 - y^2}$ pairing symmetry in the low-energy branch, or $d_{xy}$ pairing symmetry in the high-energy branch. The dashed blue lines denote the magnetic Brillouin zone boundary.}
\label{Akw_positive}
\end{figure*}

We can also investigate the low-energy excitation of the two-hole state through a momentum-space ``inverse ARPES measurement'' by injecting an electron $c_{\boldsymbol{k}\downarrow}^\dagger$ into it, which is defined as follows:
\begin{equation}\label{eq:Ap}
A^+(\boldsymbol{k},\omega)=-\mathrm{Im}\sum_n\frac{\left| _{\mathrm{1h}}\langle\Psi(n)|c_{\boldsymbol{k}\downarrow}^\dagger|\Psi_{\mathrm{G}}\rangle_{\mathrm{2h}}\right|^2}{\omega-[E_{\mathrm{1h}}(n)-E^{G}_{\mathrm{2h}}+\mu_+]+i\eta},
\end{equation}
where $E^{G}_{\mathrm{2h}}$ denotes the two-hole ground-state energy and $\mu_+$ is the chemical potential that shifts the excitation edge to zero bias. Injecting an electron annihilates one hole but leaves the associated antivortex excitation at the same position within the sudden approximation. Consequently, the energy scale differs drastically depending on the initial hole-hole distance. As shown in Fig.~\ref{Akw_positive}, both the electron-doped and hole-doped cases exhibit a two-branch dispersion, consistent with the findings in Refs.~\cite{Zhao_Weng_2024} and \cite{cancui}. The lower-energy branch corresponds to the $d_{x^2-y^2}$ Bogoliubov quasiparticles generated by breaking the two-hole Cooper-pair component, featuring nodal lines at $k_x = \pm k_y$. The higher-energy branch systematically shifts from $(\pm \pi/2, \pm \pi/2)$ to $(\pi, \pi)$ as energy increases. This branch corresponds to breaking the NNN and larger-distance pairs in the two-hole incoherent component $|\Psi_{\text{ic}}\rangle_{2\mathrm{h}}$. It reflects the intrinsic dispersion of the twisted hole $\tilde{c}_{i\sigma}$, since the leftover antivortex and the remaining hole are spatially separated after the electron injection.

The high-energy branches in the $t^{\prime} = \pm 0.25t$ cases do not differ drastically, implying that the energy scale for breaking the $\tilde{c}_{i\sigma}$ pair is nearly identical. This consistency reflects the fact that the underlying ``pairing glue'' in both the electron- and hole-doped cases is exactly the same: the pairing of $\tilde{c}_{i\sigma}$ is driven by the fusion of the single-hole incoherent components to eliminate the spin current \cite{cancui}.

However, the low-energy branch differs significantly between the two cases. In the electron-doped regime, the low-lying excitations appear at the antinodes, well separated from the nodal lines at $k_x = \pm k_y$. In contrast, for the hole-doped case, the low-energy excitations appear near the nodal lines, where the spectral weight is heavily suppressed by the $d_{x^2-y^2}$-wave pairing. The underlying reason for this difference is the same as discussed in the single-hole case: the intrinsic motion of the bare quasiparticle is significantly enhanced when $t^{\prime} > 0$, shifting the low-energy excitations into the antinodal region.

This distinct momentum-space excitation pattern has important implications for both experimental measurements and numerical simulations. Because the $d$-wave pairing order parameter changes sign under a $\pi/2$ rotation, impurity scattering that mixes these opposite-sign regions induces destructive quantum interference and breaks Cooper pairs. This effect is much more prominent in the hole-doped case, where the low-energy spectral weight is concentrated near the nodal lines, whereas the electron-doped case remains much more robust. Similarly, in finite-size numerical measurements of the Cooper pair correlation $\langle \Delta^{\dagger}_{ij} \Delta_{kl} \rangle$, a limited momentum resolution can inadvertently mix quasiparticle states with opposite signs near the nodal lines in the hole-doped case. The electron-doped case is much less susceptible to this phase cancellation because its low-energy weight is concentrated at the antinodes. This spatial separation may explain why the pair-pair correlation decays much more slowly in DMRG results for the electron-doped case~\cite{Jiang_Shengtao,Gong2021,DNSheng}. We must emphasize, however, that $\langle \Delta^{\dagger}_{kl} \Delta_{ij} \rangle$ (where $\Delta_{ij}=c_{i\uparrow}c_{j\downarrow}-c_{i\downarrow}c_{j\uparrow}$) may not directly reflect the macroscopic superfluid weight or the transition temperature, as the pairing mechanism here operates beyond conventional BCS theory, which is discussed below.

Therefore, in both the single-hole and two-hole cases, the intrinsic hopping process of the quasiparticle is enhanced beyond the resonance process between the two components in the electron-doped regime. In the following section, we will explore how this feature influences low-energy single-particle excitations in the finite-doping regime.

\section{Single-Particle Green's Function at Finite Doping}
\label{PhenoGreenFunc}
In previous sections, by analyzing the interference patterns of electron propagation in Sec.~\ref{Sec of sign structure} (with further details in Appendix~\ref{sign structure}) and combining them with VMC results in Sec.~\ref{VMC}, we have arrived at two key conclusions. On the one hand, opposite signs of $t^\prime$ induce a pronounced asymmetry between electron- and hole-doped cases, in that the electron-doped regime hosts an additional coherent propagation channel of the bare electrons. On the other hand, despite this asymmetry, our VMC results demonstrate that the ground-state wave functions in both cases share a common incoherent component $\ket{\Psi_G}$, as summarized in Eq.~\eqref{fractionalization in electron}. This component encodes strong-correlation physics and supports the observation, as emphasized in the Introduction, that electron-doped and hole-doped cuprates exhibit many similar non-Fermi-liquid features, pointing to a common underlying physical origin.

Building on these insights, we now formulate a Green’s function description of the single-particle spectrum to extend the analysis to the finite-doping regime, thereby enabling direct comparison with spectroscopic experiments such as ARPES \cite{ArmitageRMP}. The generic single-particle Green’s function corresponding to the experimentally accessible observable is denoted by $G^e$. For mathematical completeness, it can be formally defined as
\begin{equation}
G^{e}(\boldsymbol{r}_i-\boldsymbol{r}_j;\tau)
= -\langle\hat{T}_\tau c_{i\sigma}(\tau)c^\dagger_{j\sigma}(0)\rangle,
\label{Generic Ge}
\end{equation}
We emphasize that this expression only serves as a formal definition; a direct evaluation of Eq.~\eqref{Generic Ge} in terms of the ground state represented by $c$ operators is generally intractable for the strongly correlated problem of interest. Instead, we evaluate this Green’s function within the well-understood fractionalized representation shown in Eq.~\eqref{fractionalization in electron}(a) ($t^\prime<0$) and Eq.~\eqref{fractionalization in electron}(b) ($t^\prime>0$). The distinct low-energy structures give rise to single-particle excitations of fundamentally different physical character, reflected in their analytic forms and in the microscopic processes they encode.

We begin by revisiting the single-particle Green’s function in the $t^\prime<0$ regime, where the Fermi arc phenomenon can be captured by explicitly incorporating the recombination of fractionalized degrees of freedom into physical electrons. This corresponds to the resonance-driven propagation encoded in $\ket{\Psi_G}$ in Eq.~\eqref{fractionalization in electron}(a). We then turn to the $t^\prime>0$ regime, where, motivated by Eq.~\eqref{fractionalization in electron}(b), we adopt a two-fluid description in which a resonance-induced channel coexists and cooperates with an intrinsic propagation channel of the bare electrons. As we show below, the interplay between these two components naturally gives rise to the rich single-particle spectral features observed experimentally on the electron-doped side.




\subsection{Revisit: hole doped case with $t'<0$}
\label{hole dope}
Based on the ground-state wave-function structure shown in Eq.~\eqref{fractionalization in electron}, a bare hole $c$ injected into a hole-doped Mott insulator does not propagate as an elementary excitation. Instead, it rapidly fractionalizes into a vortex operator $e^{i \hat{\Omega}}$ and a composite fermion $\tilde{c}$ as introduced in Sec.~\ref{t-J fractionalization}, which constitute the incoherent component in the single-particle sector, as discussed in Sec.~\ref{VMC}. Both of these fractionalized excitations are gapped. The composite fermion $\tilde{c}$ exhibits a BCS-like gap structure, with its low-lying modes located near $(\pm \pi/2, \pm \pi/2)$, as described in Appendix~\ref{a spinon MF with t'}, while the vortex gap is controlled by the $b$-spinon RVB pairing gap~\cite{Mei_Eg,zeyuhan}. As a consequence, in the hole-doped case, the only mechanism to restore a coherent quasiparticle excitation is through the recombination of the composite fermion $\tilde{c}$ and the vortex $e^{i \hat{\Omega}}$, consistent with the VMC results discussed in Sec.~\ref{VMC}. At the mean-field level, the leading-order physical-electron Green's function $G_{0}^{\mathrm{e}}$ (where ``e'' and ``0'' denote the physical electron and the mean-field contribution, respectively) is represented by the following bubble diagram,
\begin{equation}
\begin{split}
    G_{0}^{e} (\boldsymbol{r}_i - \boldsymbol{r}_j; \tau) &= -\langle \hat{T}_\tau \tilde{c}_{i\sigma}(\tau) \tilde{c}_{j\sigma}^\dagger(0) e^{i[\hat{\Omega}_i(\tau)-\hat{\Omega}_j(0)]}\rangle_0\\
    &= D^{\tilde{c}}_{0} (\boldsymbol{r}_i - \boldsymbol{r}_j; \tau) f(\boldsymbol{r}_i - \boldsymbol{r}_j; \tau) \\
    &\equiv \diagram{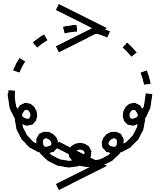}{30pt},
\end{split}
\label{Ge0}
\end{equation}
where
\begin{equation}
\begin{split}
    D^{\tilde{c}}_{0}(\boldsymbol{r}_i - \boldsymbol{r}_j;\tau) &= - \left\langle \hat{T}_\tau \tilde{c}_{i\sigma}(\tau) \tilde{c}^\dagger_{j\sigma}(0)
    e^{-i \sum_{i\rightarrow j}\phi^0_{i_s,i_s+1}}\right\rangle_0\\
    &\equiv \diagram{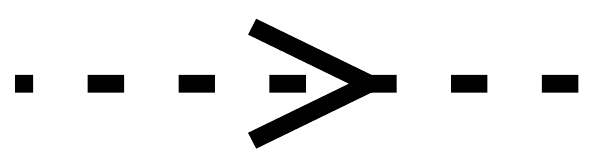}{8pt},\\
    f(\boldsymbol{r}_i - \boldsymbol{r}_j; \tau) &= \left\langle \hat{T}_\tau e^{\frac{i}{2}\Phi_i^s(\tau)} e^{-\frac{i}{2}\Phi_j^s(0)} \right\rangle_0e^{-i \boldsymbol{k}_0 \cdot (\boldsymbol{r}_i - \boldsymbol{r}_j)}\\
    &= f_0(\boldsymbol{r}_i - \boldsymbol{r}_j; \tau)e^{-i \boldsymbol{k}_0 \cdot (\boldsymbol{r}_i - \boldsymbol{r}_j)}\\
    &\equiv \diagram{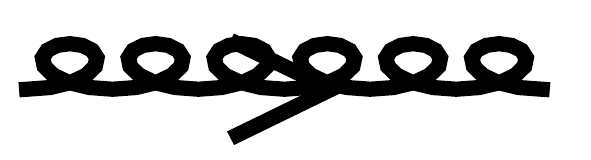}{9pt}.
\end{split}
\label{D and f}
\end{equation}
Here, $\boldsymbol{k}_0=(\pm\pi/2,\pm\pi/2)$, the phase factor $e^{-i \sum_{i \rightarrow j}\phi^0_{i_s,i_s+1}}$ is extracted from the $e^{-i(\Phi_i^0-\Phi_j^0)}$ of $e^{i[\hat{\Omega}_i(\tau)-\hat{\Omega}_j(0)]}$ to make $D_{0}^{\tilde{c}}$ gauge invariant, as detailed in Appendix~\ref{simplification of phase factor}. 
In previous works \cite{ZhangJianHao, Crossover_Fermi_arc}, the description of single-particle dynamics has been extended beyond the leading term by incorporating repeated recombination and fractionalization processes of the underlying fractionalized degrees of freedom. Within this framework, a recombined hole can further fractionalize, with the recombination (and subsequent decay) process characterized by an effective dressed vertex amplitude $\diagram{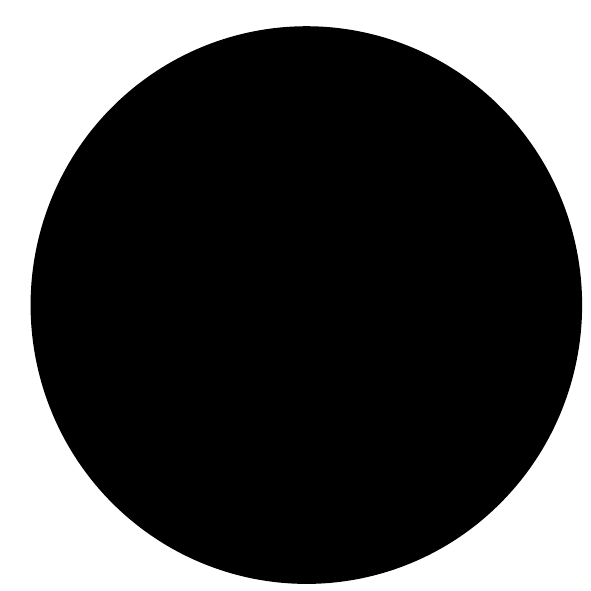}{5pt} = \lambda$. Such processes iterate to all orders and can be systematically resummed, yielding an effective description of $G^{e}$ represented by the following diagram at the RPA level:
\begin{equation}
\diagram{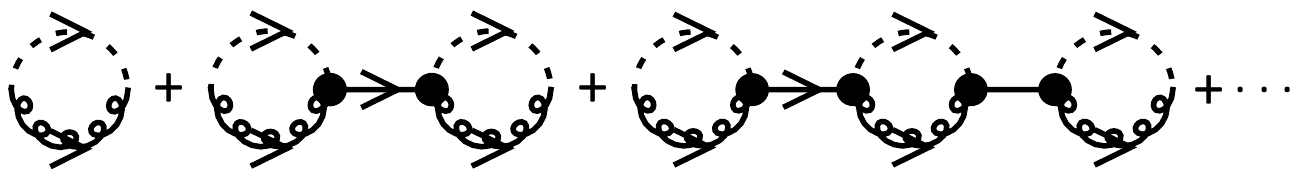}{26pt},
\label{Dyson Gctilde hole}
\end{equation}
where the solid straight line
\begin{equation}
\diagram{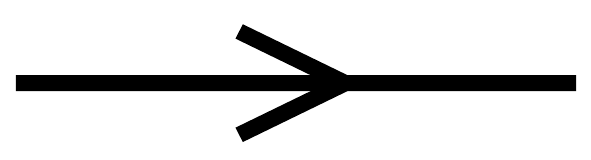}{8pt}=G^{c}_{0}(\boldsymbol{k},\omega)=\frac{1}{\omega-\epsilon_{\boldsymbol{k}}^c}
\label{Gc0}
\end{equation}
is the non-interacting propagator of the recombined electron, with a tight-binding dispersion  $\epsilon_{\boldsymbol{k}}^c=-2t(\cos k_x+\cos k_y)+4t^{\prime}\cos k_x\cos k_y+\mu_c$, where $t$ and $t^\prime$ are the bare values, $\mu_c$ is the tight-binding chemical potential of $c$. This dispersion should be distinguished from $\epsilon_{\boldsymbol{k}}^{c,\mathrm{eff}}$ in Eq.~\eqref{Et1hqp}, in which the renormalized parameters are used. The nature of the recombined hole here is worth further comment. The vortex operator $e^{i\hat{\Omega}}$ destroys the quasiparticle spectral weight \cite{Weng1997,Phasestring_Spincharge_separation}, and thereby precludes a direct observation of Landau quasiparticles in the single-particle spectrum. However, a Landau quasiparticle may still emerge as a \textit{composite mode} with a tight-binding dispersion. Since $b$-spinons form a short-range RVB state, $\Phi_i^s$ and $\Phi_j^s$ with large spatial separation are treated as uncorrelated in the long-wavelength limit, leaving $f_0(\boldsymbol{r}_i-\boldsymbol{r}_j;\tau)$ roughly constant $F_0$. Its precise value does not alter the position of spectral weight, and is chosen to be 1 in the subsequent discussion~\cite{Crossover_Fermi_arc}. Thus, the analytical expression of Eq.~\eqref{Dyson Gctilde hole} in momentum space reads
\begin{equation}
G^{e,\text{com}}(\boldsymbol{k}, \omega) = \frac{1}{{G_{0}^{e}}^{-1}(\boldsymbol{k}, \omega) - \lambda^2 G^{c}_{0}(\boldsymbol{k}, \omega)},
\label{composite mode}
\end{equation}
where the superscript ``com'' refers to the composite-mode character of the single-particle excitation appearing in the Green’s function. Physically, the structure encoded in Eq.~\eqref{Dyson Gctilde hole} and the resulting single-particle Green’s function describe precisely the \emph{resonance} motion of a doped hole discussed in Sec.~\ref{VMC}. This contribution is quantitatively captured by the kinetic-energy components $E_K^{\mathrm{ic}} + E_K^{\mathrm{cr}}$ in Eq.~\eqref{1hEt}.

To obtain a single-particle spectral weight that more closely resembles experimental observations, we further improve the bare twisted hole or composite-fermion propagator $D^{\tilde{c}}_{0}$ in Eq.~\eqref{Ge0} by incorporating self-energy corrections arising from the resonance processes discussed above. This leads to a dressed propagator $D^{\tilde{c}}=\diagram{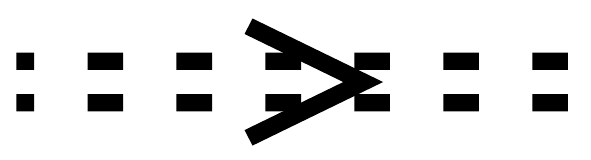}{7pt}$ represented diagrammatically by a double-dashed line, so that $G^e_0$ is replaced by $\tilde{G}^e$. The detailed derivation of this correction, along with the spectral function of $\tilde{G}^e$, can be found in Appendix~\ref{Self energy correction}. The resulting zero-energy spectral function $A^e(\boldsymbol{k}, 0)=-\mathrm{Im}\ G^{e,\text{com}}(\boldsymbol{k},0) = -\mathrm{Im} \left[ \tilde{G}_{0}^{e\ -1}(\boldsymbol{k}, 0) - \lambda^2 G^{c}_{0}(\boldsymbol{k}, 0) \right]^{-1}$. It exhibits a characteristic Fermi-arc structure, with finite spectral weight confined to the nodal region and a strong suppression in the antinodal region, in good agreement with ARPES measurements \cite{reber_origin_2012}.

\begin{figure}[h]
\centering
\includegraphics[width=0.85\linewidth]{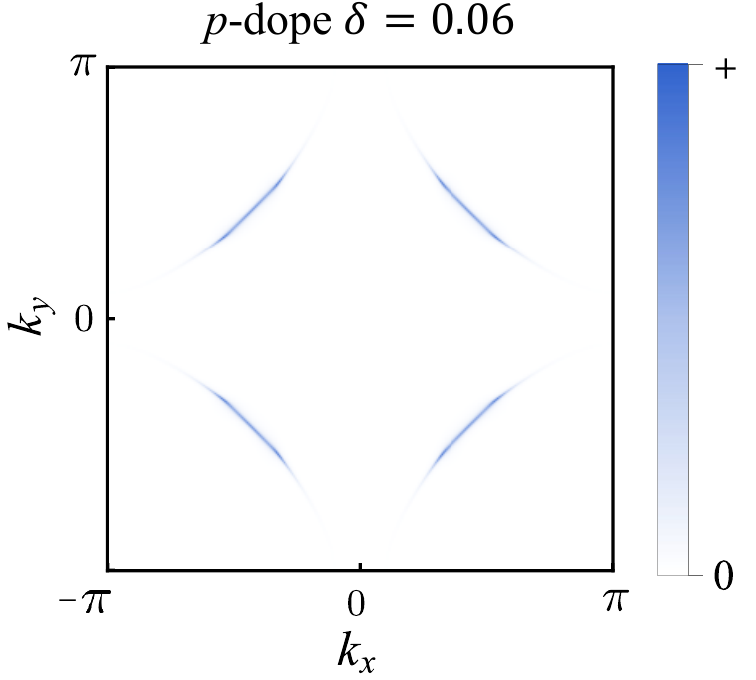}
\caption{Fermi arc spectrum at hole-doping level $\delta=0.06$ calculated using Eq.~\eqref{Dyson Gctilde hole}. The nodal spectrum here originates from the recombination of the composite fermion $\tilde{c}$ and vortex operator $e^{i\hat{\Omega}}$. Parameters corresponding to the $\delta=0.06$ row of Table~\ref{tab:typical MF values} are used, with $\lambda=0.06J$.}
\label{Fermi arc hole}
\end{figure}

\subsection{The electron-doped case with $t'>0$}
\label{Precise calculation of single particle spectrum}
As discussed in Eq.~\eqref{fractionalization in electron}, on the $t^\prime>0$ side the ground-state wave function contains, in addition to the fractionalized component $\ket{\Psi_G}$ that describes composite quasiparticle component (as discussed for the hole-doped case in Sec.~\ref{hole dope}), an extra coherent quasiparticle component $\ket{\Psi_c}$. Focusing first on the former sector $\ket{\Psi_G}$, we show in Appendix~\ref{a spinon MF with t'} that the physical nature of the composite mode, which involves the composite fermion $\tilde{c}$, is insensitive to the sign of $t'$. Its low-lying modes are always concentrated near $(\pm \pi/2,\pm \pi/2)$, which implies that the quasiparticles emerging from the RPA recombination process described by Eq.~\eqref{Dyson Gctilde hole} are likewise located in the nodal region. This conclusion is fully consistent with the VMC results presented in Sec.~\ref{VMC}, which demonstrate that quasiparticle excitations near the nodal region with Fermi arc predominantly originate from a strong resonance process with an incoherent composite component, irrespective of the sign of $t'$. 

By contrast, the coherent quasiparticle component $\ket{\Psi_c}$ exhibits a qualitatively different momentum structure: VMC calculations show that, in the electron-doped case with $t'>0$, quasiparticles arising from the intrinsic propagation channel are concentrated near the antinodal region. In the following, we first employ the Green’s function formalism to reproduce and analyze this momentum-selective behavior.

The propagation of a single electron through a series of fractionalization and recombination processes in terms of the composite fermion $\tilde{c}$ and the vortex field $e^{i\hat{\Omega}}$, can be represented in the following diagram
\begin{equation}
\diagram{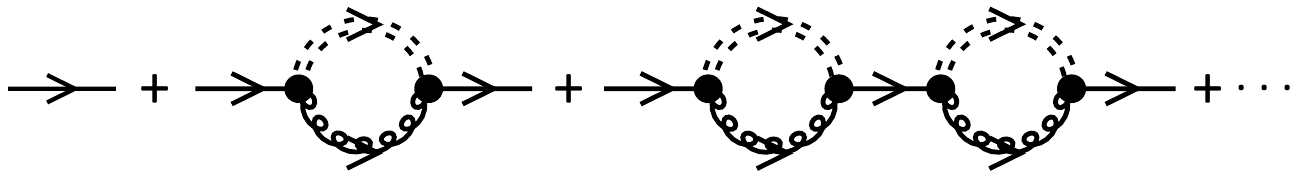}{28pt}.
\label{Dyson equation of c}
\end{equation}
The above diagram can be written in the following analytical form:
\begin{equation}
\begin{split}
    & G^{e,\text{qp}}(\boldsymbol{r}_i-\boldsymbol{r}_j;\tau)=G^{c}_{0}(\boldsymbol{r}_i-\boldsymbol{r}_j;\tau)\\
    &+\lambda^2 \int d \tau^\prime d \tau^{\prime\prime}\sum_{\boldsymbol{r}^{\prime},\boldsymbol{r}^{\prime\prime}}G^{c}_{0}(\boldsymbol{r}_i-\boldsymbol{r}^{\prime};\tau-\tau^{\prime})\\
    &\times \tilde{G}^{e}(\boldsymbol{r}^{\prime}-\boldsymbol{r}^{\prime\prime};\tau^{\prime}-\tau^{\prime\prime})G^{c}_{0}(\boldsymbol{r}^{\prime\prime}-\boldsymbol{r}_j;\tau^{\prime\prime})+\cdots,
\end{split}
\label{Ge AN}
\end{equation}
where the superscript ``qp'' denotes the coherent quasiparticle nature of the single-particle excitation appearing in the Green’s function, as opposed to the composite mode in Eq.~\eqref{composite mode}. The Green's function $\tilde{G}^e$ has a similar meaning to that in Eq.~\eqref{Ge0}, with the only difference being that $D_{0}^{\tilde{c}}$ is replaced by $D^{\tilde{c}}$, as mentioned in the previous discussion of the hole-doped scenario. In momentum and frequency space, the Dyson equation can be organized in a more compact form, which is
\begin{equation}
    G^{e,\text{qp}}(\boldsymbol{k},\omega)=\frac{1}{{G^c_0}^{-1}(\boldsymbol{k},\omega)-\lambda^2 \tilde{G}^{e}(\boldsymbol{k},\omega)}.
    \label{Gc Dyson}
\end{equation}
Although the solid line appearing in Eq.~\eqref{Dyson equation of c} is formally identical to Eq.~\eqref{Gc0}, it no longer signifies a composite mode bound by the incoherent background. Instead, it describes electrons that evade fractionalization at leading order and therefore propagate intrinsically. As shown in Appendix~\ref{Self energy correction}, the quantity $-\mathrm{Im}\,\tilde{G}^{e}(\boldsymbol{k},\omega)$, which enters the Dyson equation [Eq.~\eqref{Gc Dyson}] as the imaginary part of the electron self-energy and governs the physical electron lifetime, exhibits a momentum-dependent profile resembling a Fermi arc, with its spectral weight concentrated mostly near the nodes. As a result, the nodal quasiparticle excitations around $(\pm \pi/2,\pm \pi/2)$ in Eq.~\eqref{Gc Dyson} are heavily scattered by the $\tilde{c}$ sector. This leads to a severe suppression of coherent spectral weight at the nodes, while leaving relatively well-defined quasiparticle features intact primarily in the antinodal region.

Taken together, the intrinsic propagation channel of quasiparticles, corresponding to $E_K^{\mathrm{qp}}$ in Eq.~\eqref{1hEt} within the VMC framework and to the diagram in Eq.~\eqref{Dyson equation of c} within the Green’s function formalism, gives rise to coherent quasiparticle signatures primarily in the antinodal region, where it is largely free from scattering by the incoherent excitations associated with the twisted hole $\tilde{c}$ that are concentrated near the nodal region. By contrast, the resonance-induced propagation channel, corresponding to $E_K^{\mathrm{ic}}+E_K^{\mathrm{cr}}$ in Eq.~\eqref{1hEt} and to the diagram in Eq.~\eqref{Dyson Gctilde hole}, emerges directly from the incoherent sector and thus produces low-energy spectral weight predominantly in the nodal region. As a result, the two channels naturally lead to momentum-separated excitations with distinct physical characters in complementary momentum regions across the Brillouin zone.

This naturally motivates a two-fluid picture for the electron-doped regime. Specifically, the low-energy physics consists of two coexisting components: a coherent quasiparticle component $\ket{\Psi_c}$ with density $n_c$, which undergoes intrinsic propagation governed by Eq.~\eqref{Ge AN} and predominantly populates the antinodal region; and a composite component $\ket{\Psi_G}$ with density $n_{\tilde{c}}$, which gives rise to composite-mode excitations via the resonance processes described by Eq.~\eqref{Dyson Gctilde hole} and dominates the nodal region. Accordingly, the total carrier density obeys the sum rule $\delta = n_c + n_{\tilde{c}}$, where $\delta$ denotes the doping level. The detailed partition between $c$ and $\tilde{c}$ is determined by the delicate interplay between the energy gain originating from the resonance between both components and the intrinsic kinetic energy associated with $c$. Near the low-doping limit, $c$ primarily dominates the spectral weight because it can effectively hop via the NNN bonds. At higher doping, the resonance motion becomes increasingly prominent, ultimately establishing a two-fluid resonance regime.

\begin{figure*}
    \centering    \includegraphics[width=1\linewidth]{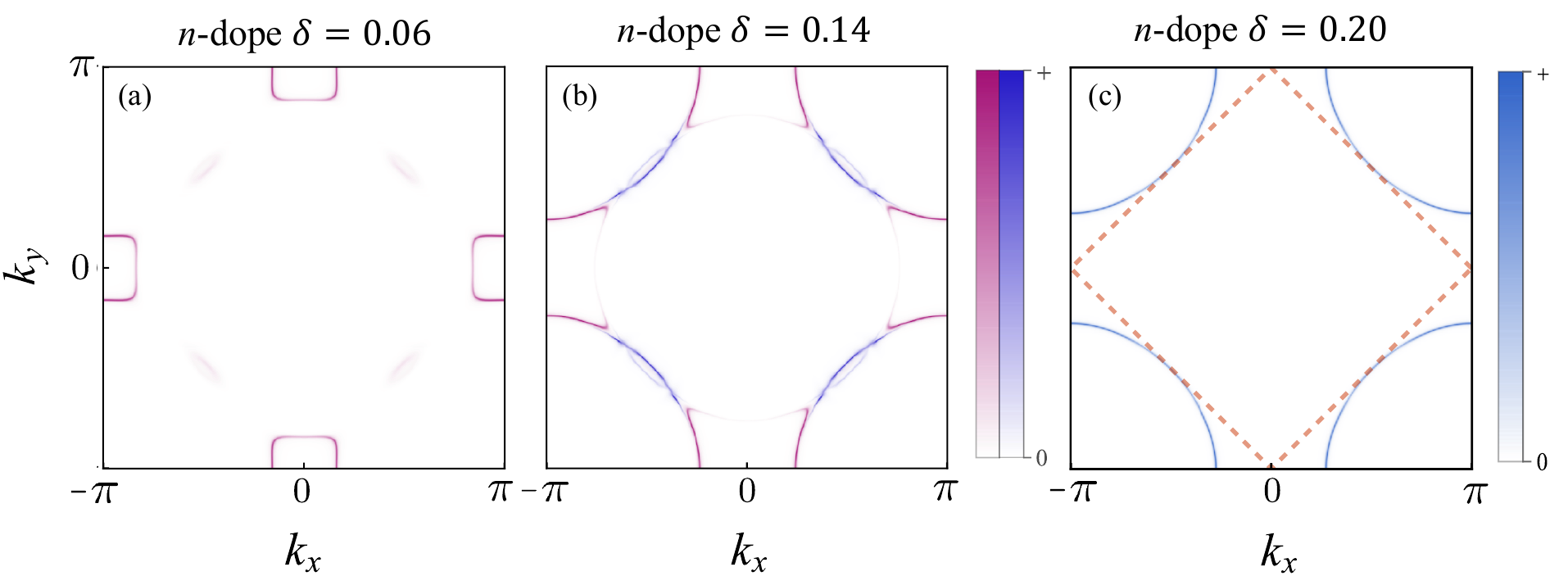}
    \caption{Zero-energy single-particle spectra at finite doping. (a, b) Spectra for doping levels $\delta = 0.06$ and $0.14$, respectively. At low doping, the coherent quasiparticle spectral weight is concentrated in the antinodal region, shown in pink in (a). With increasing doping, spectral weight emerges near the nodal region (blue), which is attributed to a composite mode originating from the underlying incoherent component. The antinodal excitations (pink) retain their coherent quasiparticle character. In this intermediate-doping regime, the two modes are well separated in momentum space. (c) High-doping case ($\delta = 0.20$), where the antinodal quasiparticle and nodal composite-mode spectral weights both grow and eventually merge to form a large Fermi surface. The orange dashed line denotes the AFM Brillouin zone boundary, which gradually moves away from the tight-binding Fermi surface as doping increases, leading to a weakened band-folding effect. For panels (a)--(c), the mean-field parameters are taken from the $\delta=0.06$, $0.14$, and $0.20$ rows of Table~\ref{tab:typical MF values}, respectively, with $J_{\mathrm{cp}} = 0.4J$, $0.25J$, and $0$. Here, $J$ is the superexchange coupling strength, and $\lambda$ is set to $0.1J$ for all three panels.}
    \label{Fermi surface}
\end{figure*}

This momentum-selective two-fluid scenario is particularly well controlled in the low- to intermediate-doping regime, where the two components remain well separated in momentum space at low energies. This key feature is confirmed by the VMC calculations in Sec.~\ref{Energy band asymmetry induced by t'}. With increasing doping, although the fundamental two-fluid description remains valid, the sharp momentum-space separation between the two components gradually weakens. This evolution is illustrated in Fig.~\ref{Fermi surface} (discussed in detail below), where the spectral weight of the Fermi arc in the nodal region progressively overlaps with the quasiparticle spectral weight in the antinodal region, blurring their momentum-space distinction. Nevertheless, the underlying two-fluid framework continues to capture the essential physics of the ground state across a wide range of finite dopings.

Notably, in agreement with experimental observations, we have argued in Sec.~\ref{Sec of sign structure} that a positive $t^\prime$ is more compatible with antiferromagnetic order, which in turn leads to enhanced antiferromagnetic fluctuations. To capture the influence of AFM fluctuations, we introduce a scattering term representing the interaction of charge carriers with a \textit{static} Néel order, $J_{\text{cp}}\sum_{i}(-1)^i c_{i\alpha}^\dagger \rho^z_{\alpha\beta} c_{i\beta}$, where $\rho^z$ is the Pauli matrix. Here, $J_{\text{cp}}$ represents the coupling strength to the AFM order. For analytical simplicity, we model the AFM order as static and long-ranged. Thus, the corresponding folded Green’s functions read
\begin{equation}
   G^{e,\ell}_{\text{fold}}(\boldsymbol{k},\omega)= \frac{1}{{G^{e,\ell}(\boldsymbol{k},\omega)}^{-1}-J_{\text{cp}}^2 G^{e,\ell}(\boldsymbol{k} + \boldsymbol{k}_{\text{AF}},\omega)}.
\label{folded equation}
\end{equation}
where $\ell=\text{com},\text{qp}$ denotes the Green's functions evaluated using Eqs.~\eqref{composite mode} and \eqref{Ge AN}, respectively. Note that the static AFM order approximation here does not alter the essential physics, as the low-energy spin-fluctuation spectral weight remains concentrated near the antiferromagnetic wave vector $\boldsymbol{k}_{\text{AF}} = (\pi,\pi)$. A more general treatment, which preserves the qualitative features and is applicable in the absence of true magnetic long-range order but in the presence of persistent short-range antiferromagnetic correlations, can be formulated within a nonlinear $\sigma$ model framework, as outlined in Appendix~\ref{App NLSM}.

We emphasize that, in the discussion above, the scattering of the physical electron $c$ arises from two independent mechanisms: one associated with the presence of $\tilde{c}$ fermions in the nodal region, and the other stemming from proximity to the AFM Brillouin-zone boundary. These two mechanisms deplete the nodal spectral weight in distinct and independent ways. The former is a purely strong-correlation effect, whereas the latter corresponds to the conventional band-folding mechanism induced by AFM order.

Next, we examine the doping dependence of the single-particle spectrum in detail, disentangling the distinct contributions from the $\tilde{c}$ and $c$ components. In the following calculations of the spectral functions, the doping-dependent mean-field parameters for the $\tilde{c}$ propagator are determined by self-consistently solving the saddle-point equations given in Appendix~\ref{Effective Hamiltonian}. These parameters uniquely fix the mean-field inputs of the composite sector, whereas the AFM folding strength $J_{\mathrm{cp}}$ and the decay rate $\lambda$ are treated as phenomenological parameters.

\subsubsection{Low doping regime}
In the low-doping regime, as verified by the previous VMC calculations in Sec.~\ref{Energy band asymmetry induced by t'}, doped holes can maximize their kinetic energy gain by moving through the next-nearest-neighbor hopping channel, thereby avoiding the frustration associated with nearest-neighbor hopping. As a consequence, the doped holes retain their quasiparticle character and appear predominantly in the antinodal region.

This observation is further supported by an analytical argument. To evaluate the antinodal prominence analytically, we turn the argument around by introducing the nodal scattering sector. Suppose that a fraction of the doped holes fractionalizes into an incoherent mode $\tilde{c}$. Based on the mean-field analysis in Appendix~\ref{Effective Hamiltonian}, the $\tilde{c}$ fermions form pockets centered around the nodal points $(\pm\pi/2,\pm\pi/2)$. The presence of these nodal pockets opens an intense, momentum-selective scattering channel, forcing the single-particle Green’s function into an antinodal-dominated form:
\begin{equation}
G^e(\boldsymbol{k},\omega)=G^{e,\mathrm{qp}}_{\mathrm{fold}}(\boldsymbol{k},\omega),
\label{low_doping}
\end{equation}
which is derived by substituting Eq.~\eqref{Dyson equation of c} into Eq.~\eqref{folded equation}. Microscopically, Eq.~\eqref{Dyson equation of c} demonstrates that the intense coupling to the nodal $\tilde{c}$ mode degrades the nodal coherence far more severely than that at the antinodes. Because this loss of coherence heavily impedes itinerant hopping, it cripples the kinetic energy gain near the nodes. Amplified by the AFM long-range order in Eq.~\eqref{folded equation}, this nodal degradation ensures that upon doping, the coherent quasiparticles seeking to maximize their kinetic energy are completely blocked from the nodes and manifest entirely within the antinodal region, as displayed in Fig.~\ref{Fermi surface}(a).

\subsubsection{Intermediate doping regime}
Upon increasing doping, a fraction of $c$ decay into $\tilde{c}$ and $e^{i\hat{\Omega}}$, forming four pockets around $(\pm\pi/2,\pm\pi/2)$. Through the same mechanism underlying the formation of the Fermi arc in Sec.~\ref{hole dope}, a composite mode emerges in the nodal regime. Thus, the experimentally observed spectrum now contains both the bare-quasiparticle contribution described by Eq.~\eqref{Dyson equation of c} and the composite mode described by Eq.~\eqref{Dyson Gctilde hole}. In the propagator of the composite mode, we replace the bare single–dashed line $D_{0}^{\tilde{c}}$ with the dressed double–dashed line $D^{\tilde{c}}$. In our calculation, we assume that only a fraction of the doped holes form $\tilde{c}$ pockets. Specifically, we take the density $n_{\tilde c}$ to be $\delta/2$ ($\delta=n_c+n_{\tilde c}$), which is merely a convenient choice and does not carry intrinsic physical significance.

After taking into account the AFM folding effect, the Green's function reads
\begin{equation}
    G^e(\boldsymbol{k},\omega)=G^{e,\text{qp}}_{\text{fold}}(\boldsymbol{k},\omega)+G^{e,\text{com}}_{\text{fold}}(\boldsymbol{k},\omega).
\label{intermediate_doping}
\end{equation}
Here, the linear combination in Eq.~\eqref{intermediate_doping} serves as a phenomenological superposition intended to illustrate the momentum-selective low-energy band structure, rather than strictly preserving the local sum rule for total spectral weight. The corresponding spectral weight is shown in Fig.~\ref{Fermi surface}(b). A clear separation of spectral weight with distinct physical origins is observed. The four pockets in the antinodal region, shown in pink, arise from the quasiparticle spectral weight $-\mathrm{Im}\,G^{e,\text{qp}}_{\textrm{fold}}$. The composite-mode spectral weight $-\mathrm{Im}\,G^{e,\text{com}}_{\textrm{fold}}$ gives rise to four blue pockets in the nodal region. This momentum-selective behavior at finite doping naturally extends the physics of the single-hole-doped limit discussed above.

\subsubsection{High doping regime}
The calculation of the single-particle spectral weight in the high-doping regime follows Eq.~\eqref{intermediate_doping}, with procedures similar to those used in the intermediate-doping regime. The resulting single-particle spectrum is shown in Fig.~\ref{Fermi surface}(c). The re-emergent large Fermi surface at high doping, however, differs from that in the intermediate-doping regime in two important aspects.

First, the Fermi-surface folding is further weakened at high doping. This suppression has two distinct origins. It arises partly because the strength of AFM fluctuations (or equivalently, the effective AFM exchange coupling) decreases with doping and partly because the underlying tight-binding Fermi surface moves farther from the AFM Brillouin-zone boundary, which further diminishes the folding effect (the AFM boundary is indicated in Fig.~\ref{Fermi surface}(c)).

Second, the formation of a large Fermi surface reflects a gradual loss of momentum-space separation between the two components. With increasing doping, the composite-mode spectral weight in the nodal region extends to the antinodal region, while quasiparticle excitations can also persist in the nodal regime due to the weakening of antiferromagnetic order and the associated reduction of frustration. Consequently, the two components progressively merge, leading to a crossover in which the momentum–component locking becomes ill-defined, even though the underlying two-fluid description might remain valid.

\section{Discussion}
\label{Discussion}
\subsection{Comparison of Phase-String Theory with Other Theoretical Models}
It is instructive to compare the present phase-string–based description with several existing theoretical frameworks that have been proposed to account for the pseudogap electronic structure in cuprates.

One of the earliest phenomenological constructions is the Yang–Rice–Zhang (YRZ) Green’s function~\cite{YRZ}. In this approach, the Green’s function is built upon a doped RVB spin liquid, where the doped holes scatter off the background spinon Fermi surface. This scattering reconstructs the large Fermi surface into small pockets. However, the spectral weight on the backside of the pocket is strongly suppressed, such that the experimentally observable structure appears as Fermi arcs. While this construction captures several phenomenological features of the pseudogap regime, the existence of the pocket structure is still under debate.

A different route is provided by the ancilla model~\cite{Ancilla_model,Ancilla_model1}, which reformulates the single-band Hubbard model into an effective three-layer structure consisting of an itinerant electron layer coupled to two layers of localized spins via interlayer superexchange. Within this framework, the physical electron operator remains intact, while the spin sector is enlarged by auxiliary degrees of freedom, thereby endowing the theory with a multiband character similar to that of the Kondo lattice model. The electron consequently persists as a well-defined low-energy quasiparticle whose propagator is renormalized through scattering off the spin fluctuations~\cite{Christos2024}. In this picture, the electron–hole asymmetry originates from the dispersion asymmetry of the spinon Fermi surface. Nevertheless, the emergence of such asymmetry fundamentally relies on the introduction of extra auxiliary degrees of freedom beyond the original single-band Hilbert space.

By contrast, the present phase-string formulation of the single-band $t$-$t^\prime$-$J$ model provides an intrinsically single-band description. The two-component electronic structure emerging in the electron-doped regime arises entirely from the internal sign structure of the single-band problem rather than from a phenomenological Green’s function or an explicit enlargement of the Hilbert space. In particular, the sign of $t^\prime$ qualitatively modifies both the ground state (see Eq.~\eqref{fractionalization in electron}) and the nature of the single-particle excitations. From this perspective, the electron–hole asymmetry observed experimentally can be understood as a direct consequence of the phase-string sign structure. Our theory supports an intrinsic Fermi-arc structure rather than pocket physics.

Here, we briefly note that quantum oscillation signals~\cite{Kartsovnik_2011,Higgins_2018,Helm2009,Doiron-Leyraud2007,Barisic2013} are often interpreted as direct evidence for the existence of Fermi pockets. Nevertheless, because such experiments are performed under high magnetic fields, the BCS pairing of $\tilde{c}$ is destroyed. As a consequence, $\tilde{c}$ forms four Fermi pockets at zero energy that can be directly detected experimentally, as discussed in detail in Refs.~\cite{Ma_2014,Crossover_Fermi_arc}. This $\tilde{c}$-related pocket structure is fundamentally different from the one shown in Refs.~\cite{YRZ,Christos2024}, which manifests exclusively in the single-particle spectrum.

\subsection{Superconductivity and spin resonance mode}
In cuprate superconductors, a universal scaling law, $E_g \approx 6T_c$, linking the resonance energy $E_g$ of the spin-fluctuation spectrum to the superconducting transition temperature $T_c$, has been widely established by neutron scattering~\cite{Y123_Eg,Bi2212_Eg,Tl2201_Eg,Wilson2006,NCCO_Eg,Yu2009} and Raman spectroscopy~\cite{Bi2212_Eg1,Hg1223_Eg,La214_Eg,Stadlober_Eg,DevereauxHackl2007}. This robust relation holds across a diverse family of compounds, including both hole-doped~\cite{Y123_Eg,Bi2212_Eg,Bi2212_Eg1,Hg1223_Eg,Tl2201_Eg,La214_Eg} and electron-doped~\cite{Stadlober_Eg,Wilson2006,NCCO_Eg} systems.

Within our phase-string framework, the superconducting transition has been systematically investigated via a renormalization-group approach~\cite{Mei_Eg,Ma_2014}, providing a microscopic foundation for this empirical scaling. In this context, the universal scaling between $E_g$ and $T_c$ firmly supports the presence of the $\ket{\Phi_b}$ component within $\ket{\Psi_G}$ of Eq.~\eqref{fractionalization in electron} for both signs of $t^\prime$. In particular, thermally excited $b$-spinon vortices act to disorder the phase of the superconducting condensate formed by $\tilde{c}$ pairs. The proliferation of these topological excitations destroys long-range phase coherence, thereby dictating the scale of $T_c$. At zero temperature, the coupling between spin and charge degrees of freedom further renormalizes the superfluid stiffness from its bare, doping-dependent value down to a reduced scale that tracks $T_c$, or equivalently, $E_g$~\cite{zeyuhan,PengYe}.

Moreover, the experimental gap-to-$T_c$ ratio is maintained at an anomalously large value of $2\Delta/k_{\mathrm{B}}T_c \approx 7.5$~\cite{Niestemski2007}, well beyond the conventional weak-coupling BCS limit. Taken together, the interlinked energy scales collectively signal a strong-coupling origin of the superconductivity. The robust connection among these energy scales across both doping regimes underscores a unified underlying structure encoded in $\ket{\Psi_G}$ of Eq.~\eqref{fractionalization in electron}. This realization renders the standard picture of AFM fluctuation-mediated pairing an inadequate description of cuprate superconductivity.

\subsection{Dispersion Anisotropy in the Superconducting State}
At finite doping, in the superconducting state, excitations formed by composite modes remain low in energy along the nodal direction, whereas antinodal quasiparticle states lie at much higher energies due to the fully opened $d$-wave gap. Consequently, the relevant low-energy Green's function is obtained via Eq.~\eqref{composite mode}, where phase coherence is manifested by a non-vanishing expectation value $\braket{e^{i\Phi_i^s}}\neq 0$. To incorporate the pairing dynamics, this Green's function is cast into the Nambu matrix form, based on the Nambu spinor
$\Psi_{\tilde{c}}(\boldsymbol{k})=(\tilde{c}_{\boldsymbol{k}\uparrow}\ \ \tilde{c}_{-\boldsymbol{k}\downarrow}^{\dagger})^T$ and $\Psi_{c}(\boldsymbol{k})=(c_{\boldsymbol{k}\uparrow}\ \ c_{-\boldsymbol{k}\downarrow}^{\dagger})^T$. The non-interacting Green's functions of $\tilde{c}$ and $c$ are given below and are also discussed in Appendix~\ref{Self energy correction}:
\begin{equation}
    \begin{split}
        \hat{D}^{\tilde{c}}_0(\boldsymbol{k},\omega)&=\frac{1}{\omega\tau_0+\xi_{\boldsymbol{k}}\rho_z-\Delta_{\boldsymbol{k}}\rho_x},\\
        G^{c}_{0,\text{SC}}(\boldsymbol{k},\omega)&=\frac{1}{\omega\tau_0-\epsilon_{\boldsymbol{k}}^c\rho_z-\Delta_{\boldsymbol{k}}^c\rho_y},
    \end{split}
\end{equation}
with Pauli matrices $\rho_x$ and $\rho_y$ in the denominator reflecting the choice of the relative pairing phase in order to gap out most of the Fermi surface and optimize the energy gain~\cite{Crossover_Fermi_arc}. Here, $\Delta_{\boldsymbol{k}}^c = J_{\text{eff}}\Delta^a(\cos k_x - \cos k_y)$ denotes the $d$-wave pairing order parameter of the quasiparticles, where $J_{\text{eff}}$ is the effective superexchange coupling, and $\Delta_{\boldsymbol{k}} = 2\gamma\Delta^a\sqrt{\cos^2 k_x + \cos^2 k_y}$ is the pairing amplitude of $\tilde{c}$. $J_\text{eff}$, $\gamma$ and $\Delta^a$ are the parameters of the mean-field calculation defined in Appendix~\ref{Effective Hamiltonian}. Moreover, the propagator of the phase factor $(e^{\frac{i}{2}\Phi_i^s(\tau)}\ \ e^{-\frac{i}{2}\Phi_i^s(\tau)})^T$ is
\begin{equation}
    \hat{f}(\boldsymbol{r}_i-\boldsymbol{r}_j,\tau)=[F_0\tau_0+F_0\braket{e^{i\Phi_i^s(\tau)}}\tau_x] e^{-i \boldsymbol{k}_0 \cdot (\boldsymbol{r}_i - \boldsymbol{r}_j)}.
\end{equation}

\begin{figure}
    \centering
    \includegraphics[width=0.85\linewidth]{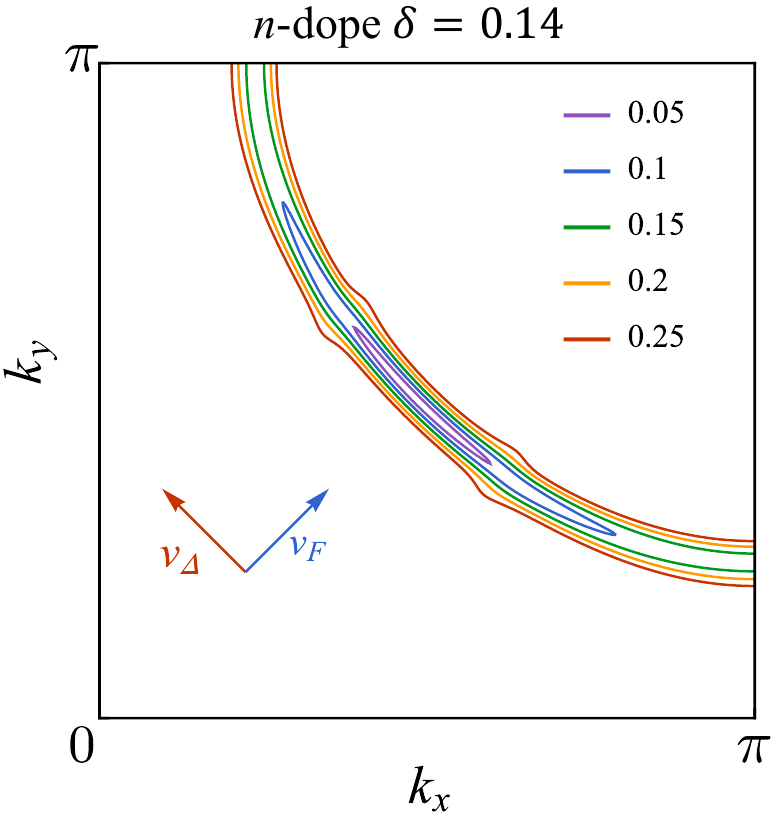}
    \caption{Constant-energy contours ($\omega = 0.05$ to $0.25$) of the pole positions derived from the superconducting Green's function in the first quadrant of the Brillouin zone for the electron-doped case. The dispersions near the Dirac node along the directions indicated by the blue and red arrows yield the Fermi velocity $v_F$ and gap velocity $v_\Delta$, respectively, which can also be inferred from the spacing of the contour lines. A pronounced anisotropy with $v_F \gg v_\Delta$ is clearly evident. Parameters corresponding to the $\delta=0.14$ row of Table~\ref{tab:typical MF values} are used, with $\lambda=0.1J$.}
    \label{contour SC poles}
\end{figure}
Combining the discussion above, the single-particle Green's function of the composite mode in the superconducting phase can be expressed according to Eq.~\eqref{composite mode}:
\begin{equation}
G^{e,\text{com}}_{\text{SC}}(\boldsymbol{k},\omega)=\frac{M_0I+\boldsymbol{M}\cdot\boldsymbol{\rho}}{(\omega^2-E_+^2)(\omega^2-E_-^2)},
\end{equation}
where $I$ is the identity matrix, $\boldsymbol{\rho}=(\rho_x, \rho_y, \rho_z)$. $M_0$ and $\boldsymbol{M}$ combine to a four-dimensional vector whose complicated expression does not affect the pole positions. The pole positions are determined by the zeros of the denominator, given by $E_{\pm}=\sqrt{A/2\pm \sqrt{B}/2}$ with $A=(\Delta_{\boldsymbol{k}+\boldsymbol{k}_0})^2+(\Delta_{\boldsymbol{k}}^c)^2+(\xi_{\boldsymbol{k}+\boldsymbol{k}_0})^2+(\epsilon_{\boldsymbol{k}}^c)^2+2\lambda^2$ and $B=[(\Delta_{\boldsymbol{k}+\boldsymbol{k}_0})^2-(\Delta_{\boldsymbol{k}}^c)^2+(\xi_{\boldsymbol{k}+\boldsymbol{k}_0})^2-(\epsilon_{\boldsymbol{k}}^c)^2]^2+4\lambda^2[(\Delta_{\boldsymbol{k}+\boldsymbol{k}_0})^2+(\Delta_{\boldsymbol{k}}^c)^2+(\xi_{\boldsymbol{k}+\boldsymbol{k}_0}-\epsilon_{\boldsymbol{k}}^c)^2]$. Figure~\ref{contour SC poles} shows constant-energy contours of the lower branch $E_{-}$ in the first quadrant, highlighting a strong velocity anisotropy ($v_F \gg v_\Delta$) near the Dirac cone. Band-folding effects are omitted in the analytical evaluation for simplicity. While incorporating band folding complicates the algebraic expressions by shifting the pole locations, it leaves the dispersion anisotropy intact. Neglecting this effect thus yields tractable analytical results while preserving our core conclusion.

\section{Conclusion}
In this paper, we have presented a microscopic framework that connects the particle--hole asymmetry observed in the single-particle spectra of cuprate superconductors to the underlying Mott physics.
In particular, the two questions posed in the Introduction have been resolved based on the $t$-$t^\prime$-$J$ model as follows. 

Question (a), addressing the origin of the asymmetry associated with the sign of $t^\prime$, is directly answered by our VMC calculations of the single-hole-doped $t$–$t^\prime$–$J$ model. For $t^\prime < 0$, the low-energy manifold $\ket{\Psi_G}$ exhibits a two-component structure consisting of a bare quasiparticle component $c$ and an incoherent fractionalized component $\tilde{c}$. Here, the kinetic energy gain is dominated by a strong resonance between these two components, forming a ``cat state''. When $t^\prime/t > 0.09$, in addition to this resonating structure around the nodal region, we identify a new source of kinetic energy gain arising solely from an emerging component, namely an independent quasiparticle mode (band) near the antinodal region. Consequently, the low-energy manifold transitions to $\ket{\Psi_c}\otimes\ket{\Psi_G}$, which naturally captures this momentum-space dichotomy.

Question (b) concerns the electron–hole asymmetry observed at finite doping. Building on the single-hole-doped case, a two-fluid description naturally emerges on the electron-doped side ($t^\prime=0.25t$). At finite doping, the single-particle Green’s function provides an additional perspective on the nature of low-energy quasiparticle excitations. In particular, a pronounced momentum-space dichotomy develops. The nodal spectrum arises as a composite mode from the recombination of $\tilde{c}$ and $e^{i\hat{\Omega}}$ and shares the same physical origin as the Fermi arcs in the $t^\prime < 0$ regime, whereas the antinodal spectrum originates solely from an additional coherent quasiparticle component.

Taken together, our responses to questions (a) and (b) provide a coherent picture for understanding the electron–hole asymmetry in doped Mott insulators. Namely, the underlying statistical sign structure changes from conventional Fermi statistics to the phase-string sign structure upon the opening of the Mott gap in the $t$-$J$ model. By adding the NNN hopping $t^\prime$, the dopants can access another hopping channel (diagonal hopping) to partially circumvent the phase-string frustration associated with the NN hopping. On the $t^\prime>0$ side, this process can become important through the emergence of a new antinodal band in the low-energy sector, which significantly influences the low-doping phases and may explain the experimental observations in electron-doped cuprates.

The weakened non-Fermi liquid behavior observed in electron-doped cuprates arises naturally within our two-fluid framework, as the reduced weight of the $\tilde{c}$ component reflects a reduced, yet still non-negligible, correlation effect. To be more specific, the resonance mode of $E_g$ in both electron- and hole-doped cuprates as well as its scaling with $T_c$ are explained comprehensively in Ref. \cite{Mei_Eg}, in which a renormalization-group calculation relates the loss of superconducting phase coherence to the proliferation of $b$-spinon vortices, whose fugacity is determined by $E_g$. The fact that $E_g \approx 6T_c$ holds in both doping regimes indicates that the superconducting transition is controlled by $E_g$, an energy scale arising from the local moments. This interpretation goes beyond a weak-coupling description based on itinerant degrees of freedom \cite{Hourglass}. The thermoelectric properties are discussed in Ref.~\cite{Thermal_Hall}, where the edge states of $b$-spinon contribute to the Hall transport. Strange-metal behavior is also discussed in Ref. \cite{Strange_metal}, where it arises from scattering between $\tilde{c}$ and random flux generated by disordered $b$-spinon vortices.

Furthermore, the pairing amplitude $\braket{cc}$ obtained in DMRG calculations may be inadequate to faithfully reflect the true superconducting transition temperature. The results of Ref.~\cite{Jiang_Shengtao} show a significantly stronger pairing in the doped $t$-$t^\prime$-$J$ model for $t^\prime>0$ than for $t^\prime<0$.  This enhancement can be attributed to the presence of an independent quasiparticle component in the $t^\prime > 0$ ground state. As demonstrated by our two-hole VMC results in Sec.~\ref{Two-hole VMC}, the nearest-neighbor Cooper pair for $t^\prime > 0$ can propagate coherently through the next-nearest-neighbor hopping channel $H_{t^\prime}$, which boosts its amplitude and enhances local pair--pair correlations. However, because the macroscopic $T_c$ is governed by the spin gap $E_g$ rather than the conventional BCS pairing amplitude, this local enhancement may not yield a higher $T_c$. Instead, our sign-structure analysis reveals that the local NN and NNN hopping processes for $t^\prime > 0$ interfere constructively within the local AFM background. This drastically reduces magnetic frustration and sustains stronger AFM correlations, thereby suppressing the spin gap $E_g$ and ultimately resulting in the lower $T_c$ observed in electron-doped cuprates.

To conclude, our investigation points to a broader scope for understanding strongly correlated physics. By revealing an emergent effective two-component structure within the $t$-$t^\prime$-$J$ model, our results suggest that rich multi-fluid behavior could arise purely from a single-band model when strong correlations are considered, without invoking explicit orbital degrees of freedom. This perspective offers a unifying lens through which seemingly disparate phenomena, including electron–hole asymmetry and unconventional normal-state behavior, may be understood, and it opens a promising avenue toward a more universal description of correlated quantum matter.

\begin{acknowledgments}
{\it Acknowledgments.---}
We acknowledge stimulating discussions with Nigel Hussey, Ya-hui Zhang, Wei Ku, Jianda Wu, Zhi-Jian Song, Shuai.A Chen, Jing-Yu Zhao and Jiahao Yang. We acknowledge financial support from MOST of China (Grant No. 2021YFA1402101) and the NSF of China (Grant No. 12347107). J.X.Z. was funded by the European Research Council (ERC) under the European Union's Horizon 2020 research and innovation program (Grant Agreement No. 853116, acronym TRANSPORT), and was also supported in part by NSF Grant No. PHY-2309135 to the Kavli Institute for Theoretical Physics (KITP).
\end{acknowledgments}

\bibliography{main}

\appendix


\renewcommand\thefigure{\thesection S\arabic{figure}}
\renewcommand\theequation{\thesection S\arabic{equation}}

\begin{widetext}

\section{Sign Structure of the $t$-$t^\prime$-$J$ Model}
\label{sign structure}
In this Appendix, we provide further discussion on the sign structure to supplement Sec.~\ref{Sec of sign structure}. As briefly noted in the main text, the $t^\prime>0$ regime is more compatible with the AFM background than its $t^\prime<0$ counterpart. This behavior can be rigorously established through a detailed analysis of the underlying sign structure. To explicitly incorporate the Marshall sign, we perform the unitary transformation $c_{i\sigma}\rightarrow(-\sigma)^i c_{i\sigma}$, under which the $t$–$t^\prime$–$J$ Hamiltonian reads
\begin{equation}
    H_{t\text{-}t^\prime\text{-}J}=t(P_{o\uparrow}-P_{o\downarrow})-t^\prime T_o-\frac{J}{2}(P_{\uparrow\downarrow}+Q),
\end{equation}
where
\begin{equation}
    P_{o\sigma}=\sum_{\langle ij\rangle}c^\dagger_{i\sigma}c_{j\sigma}+\mathrm{H.c.},
\end{equation}
\begin{equation}
    T_o=\sum_{\langle \langle ij\rangle\rangle\sigma}c^\dagger_{i\sigma}c_{j\sigma}+\mathrm{H.c.},
\end{equation}
\begin{equation}
    Q=\sum_{\langle ij\rangle}(n_{i\uparrow}n_{j\downarrow}+n_{i\downarrow}n_{j\uparrow}),
\end{equation}
\begin{equation}
    P_{\uparrow\downarrow}=\sum_{\langle ij\rangle}(c^\dagger_{i\uparrow}c_{i\downarrow}c^\dagger_{j\downarrow}c_{j\uparrow}+c^\dagger_{i\downarrow}c_{i\uparrow}c^\dagger_{j\uparrow}c_{j\downarrow}).
\end{equation}
Here, $P_{o\sigma}$ denotes the term representing the exchange of a hole and a spin-$\sigma$ via the nearest-neighbor (NN) channel, while $T_o$ represents the corresponding process through the next-nearest-neighbor (NNN) channel. $Q$ characterizes the longitudinal interaction between NN spins, and $P_{\uparrow\downarrow}$ denotes the NN spin-flip operator. The high-temperature expansion of the partition function consequently reads
\begin{equation}   
Z_{t\text{-}t^\prime\text{-}J}=\mathrm{Tr}\ e^{-\beta H_{t\text{-}t^\prime\text{-}J}}=\sum_{n=0}^{\infty} \frac{\beta^n}{n!}\mathrm{Tr}(-H_{t\text{-}t^\prime\text{-}J})^n=\sum_{n=0}^{\infty}\sum_{\{\alpha_i\}}\frac{\beta^n}{n!}\prod_{k=0}^{n-1}\bra{\alpha_{k+1}}(-H_{t\text{-}t^\prime\text{-}J})\ket{\alpha_k}.
\end{equation}

\begin{equation}
\begin{split}
&Z_{t\text{-}t^\prime\text{-}J}
= \mathrm{Tr}\!\left( e^{-\beta H_{t\text{-}t^{\prime}\text{-}J}} \right)
= \sum_{n} \frac{\beta^{n}}{n!}\, \mathrm{Tr}\!\left[ (-H_{t\text{-}t^{\prime}\text{-}J})^{n} \right]\\
&= \sum_{n=0}^{\infty} \frac{(J\beta/2)^{n}}{n!}\;
\mathrm{Tr} \sum_{C}
[\cdots
\left( -\frac{2t}{J} P_{o\uparrow} \right)
\cdots P_{\uparrow\downarrow}\cdots
\left(\frac{2t}{J} P_{o\downarrow} \right)
\cdots Q\cdots
\left( \frac{2t'}{J} T_{o} \right)],
\end{split}
\end{equation}
where $\sum_C$ denotes the sum over all expansions with $n$ terms.
We can insert a complete set of doped Ising basis states into the expansion,
\begin{equation}
    \sum_{\phi,\{l_h\}}\ket{\phi,\{l_h\}}\bra{\phi,\{l_h\}}=1,
\end{equation}
where $\{l_h\}$ denotes the locations of the holes. The half-filled and doped Ising basis states are defined as
\begin{equation}
\ket{\phi}=c^\dagger_{\sigma_1}c^\dagger_{\sigma_2}\cdots c^\dagger_{\sigma_N}\ket{\text{vec}},
\end{equation}
\begin{equation}
    \ket{\phi,\{l_h\}}=c_{\sigma l_{h1}}c_{\sigma l_{h2}}\cdots\ket{\phi}.
\end{equation}
One can easily prove that the matrix elements of $Q$ and $P_{\uparrow\downarrow}$ in the Ising basis are non-negative. It also follows easily that the matrix elements of $P_{o\sigma}$ and $T_{o}$ in the Ising basis are $-1$ or $0$. The partition function reads
\begin{equation}
    Z_{t\text{-}t^\prime\text{-}J}=\sum_{\{C\}}\tau_C W[C],
\end{equation}
where
\begin{equation}
    \tau_C=(-1)^{N^h_\downarrow[C]+N^h_{ex}[C]}[-\text{sgn}(t^\prime)]^{M_{t^\prime}[C]},
\end{equation}
and the non-negative weight $W[C]$ is
\begin{equation}
    W[C]=\sum_n \frac{(J\beta/2)^n}{n!}\left(\frac{2t}{J}\right)^{M_{t}[C]}\left(\frac{2|t^\prime|}{J}\right)^{M_{t^\prime}[C]}\delta_{n,M_{t}[C]+M_{t^\prime}[C]+M_Q[C]+M_{\uparrow\downarrow}[C]}.
\end{equation}
Here, $M_{t}[C]$, $M_{t^\prime}[C]$, $M_{Q}[C]$, and $M_{\uparrow\downarrow}[C]$ denote, respectively, the numbers of $P_{o\sigma}$, $T_o$, $Q$, and $P_{\uparrow\downarrow}$ appearing in the loop $C$. $N^h_\downarrow[C]$ and $N_{ex}^h[C]$ count the numbers of hole–down-spin exchanges and hole–hole exchanges, respectively.

For Hamiltonians with positive and negative $t^\prime$, the closed loops being summed over are identical, and so are the corresponding weights $W[C]$ for each configuration $C$. The only difference lies in the sign factor $\tau_C$, which controls the resulting interference pattern, as illustrated in the main text.

\end{widetext}

\section{Effective Hamiltonian}
\label{Effective Hamiltonian}
$\tilde{c}_{i\sigma}$ can be further decomposed as $\tilde{c}_{i\sigma}=h_i^\dagger a_{i\bar{\sigma}}^\dagger$.
Following Refs.~\cite{Weng_2011,Ma_2014}, the mean-field effective Hamiltonian $H=H_h+H_a+H_b$ is written as
\begin{equation}
\begin{split}
H_h&=-t_h \sum_{\braket{ij}}h_i^\dagger h_j e^{i (A^s_{ij}+q_e A^e_{ij})}+\text{H.c.}\\
&+\mu_h \left( \sum_{i} h_{i}^\dagger h_{i} - \delta N \right),
\end{split}
\end{equation}
\begin{equation}
\begin{split}
H_a = &-t_a \sum_{\langle ij \rangle \sigma}a_{i\sigma}^\dagger a_{j\sigma}e^{-i\phi_{ij}^0}+ \text{H.c.} - \gamma \sum_{\langle ij \rangle} \hat{\Delta}_{ij}^{a\ \dagger}\hat{\Delta}_{ij}^{a}\\
&+ \mu_a \left( \sum_{i\sigma} a_{i\sigma}^\dagger a_{i\sigma} - \delta N \right),
\end{split}
\end{equation}
\begin{equation}
    H_b=-J_s \sum_{\braket{ij}}(\hat{\Delta}_{ij}^b+\text{H.c.})+\mu_b\sum_{i\sigma}(b^\dagger_{i\sigma}b_{i\sigma}-N).
\end{equation}
Here, $J_s=\frac{J_{\text{eff}}\Delta^b}{2}$, $J_{\text{eff}}=J(1-\delta)^2-2\gamma\delta^2$, $\hat{\Delta}^a_{ij}=a_{i\uparrow}a_{j\downarrow}-a_{i\downarrow}a_{j\uparrow}$, and $\hat{\Delta}_{ij}^b=\sum_\sigma b_{i\sigma}b_{j\bar\sigma}e^{-i\sigma A^h_{ij}}$. The parameter $\gamma$ is the Lagrange multiplier that enforces $\hat{\Delta}_{ij}^{a\ \dagger}\hat{\Delta}_{ij}^{a}=\delta^2\hat{\Delta}_{ij}^{b\ \dagger}\hat{\Delta}_{ij}^{b}$, which originates from the constraint $\boldsymbol{S}_i^a=-n_i^a\boldsymbol{S}_i^b$. Here, $N$ is the number of lattice sites, and $A^h_{ij}$, $A^s_{ij}$, and $A^e_{ij}$ are the gauge fields generated by the holons, $b$-spinons, and external sources, respectively, with
\begin{equation}
\begin{split}
    A^h_{ij}&=\frac{1}{2}\sum_{l\neq i,j}[\theta_i(l)-\theta_j(l)]n_l^h,\\
    A^s_{ij}&=\frac{1}{2}\sum_{l\neq i,j}[\theta_i(l)-\theta_j(l)]\sum_\sigma \sigma n_{l\sigma}^b.
\end{split}
\end{equation}
In the following, we sketch the mean-field calculation for the fractionalization formulation introduced in the main text, following Ref.~\cite{Ma_2014}. Since the holons form a condensate and each holon carries $\pi$ flux, $A_{ij}^h$ can be viewed as a uniform flux of $\delta\pi$ per plaquette. The free energy of the $b$-spinon reads
\begin{equation}
\begin{split}
    F_b&=\sum_m[E^b_m+\frac{2}{\beta}\ln(1-e^{-\beta E^b_m})]\\
    &-2\mu_b N+J_{\text{eff}}N (\Delta^b)^2.
\end{split}
\end{equation}
Here, $E_m^b$ is the $m$-th Landau-level-like $b$-spinon energy under a $\delta\pi$ flux per plaquette, and $\Delta^b=\langle \hat{\Delta}^b_{ij} \rangle$ denotes the mean-field expectation value of the spinon pairing operator. For the $a$-spinon, we further decompose its interaction term into the pairing channel $\Delta^a=\braket{e^{-i\phi_{ij}^0}\sum_\sigma \sigma\ a_{i\sigma}^\dagger a_{j\bar{\sigma}}^\dagger}=(\Delta^a)^*$ and the hopping channel $\chi^a=\braket{e^{-i\phi_{ij}^0}\sum_\sigma a^\dagger_{i\sigma}a_{j\sigma}}=(\chi^a)^*$, so that its free energy can be written as
\begin{equation}
\begin{split}
F_a &= -\frac{2}{\beta} \sump_{\boldsymbol{k}, m=\pm} \ln \left( 2 \cosh \frac{\beta E_{\boldsymbol{k}m}}{2} \right)\\
&+2N \cdot \gamma (|\chi^a|^2 + |\Delta^a|^2) + \mu_a(1 - \delta)N.
\label{F_a}
\end{split}
\end{equation}
In the above equations, $\sump$ in $F_a$ denotes a sum over half of the Brillouin zone due to the background $\pi$ flux. In Eq.~\eqref{F_a}, $E_{\boldsymbol{k}\pm}=\sqrt{\xi_{\boldsymbol{k},\pm}^2+\Delta_{\boldsymbol{k}}^2}$, where $\xi_{\boldsymbol{k},\pm}=\pm2(t_a+\gamma \chi^a)\sqrt{\cos^2 k_x+\cos^2 k_y}+\mu_a$, is the $a$-spinon dispersion under the $\pi$ flux, which contains four Dirac cones.  $\Delta_{\boldsymbol{k}}=2\gamma\Delta^a\sqrt{\cos^2 k_x+\cos^2 k_y}$ is the $s$-wave pairing order parameter of the $a$-spinons.

The saddle-point equations $\frac{\partial (F_a+F_b)}{\partial \mu_b}=\frac{\partial (F_a+F_b)}{\partial \Delta^{b}}=0$ give the self-consistent equations for the $b$-spinon parameters $\mu_b$ and $\Delta^{b}$,
\begin{equation}
\begin{split}
\sum_{m} \frac{\lambda_{b} \coth \left(\frac{1}{2} \beta E_{m}^{b}\right)}{E_{m}^{b}} &= 2N,\\
\sum_{m} \frac{(\xi_{m}^{b})^2 \coth \left(\frac{1}{2} \beta E_{m}^{b}\right)}{E_{m}^{b}} &= 2N (\Delta^{b})^2 J_{\text{eff}},
\label{MF equ of b}
\end{split}
\end{equation}
where $\xi_m^b$ is the corresponding spectrum of the Harper matrix.

The self-consistent equations for the $a$-spinon can then be derived by solving the saddle-point equations $\frac{\partial (F_a+F_b)}{\partial \mu_a}=\frac{\partial (F_a+F_b)}{\partial \Delta^a}=\frac{\partial (F_a+F_b)}{\partial \chi^a}=\frac{\partial (F_a+F_b)}{\partial \gamma}=0$,
\begin{equation}
\begin{split}
\sump_{\boldsymbol{k},m=\pm}B_{\boldsymbol{k}m}\xi_{\boldsymbol{k}m}&=(1-\delta)N,\\
\gamma\sump_{\boldsymbol{k},m=\pm}A_{\boldsymbol{k}} B_{\boldsymbol{k}m}&=N,\\
\sump_{\boldsymbol{k},m=\pm}(-1)^{m}\sqrt{A_{\boldsymbol{k}}}B_{\boldsymbol{k}m}&\xi_{\boldsymbol{k}m}=2\chi^a N,\\
(\chi^a)^2+(\Delta^a)^2&=\delta^2(\Delta^{b})^2,\\
\label{MF equ of a}
\end{split}
\end{equation}
where $A_{\boldsymbol{k}}=\cos^2k_x+\cos^2k_y$, $B_{\boldsymbol{k}m}=\frac{\tanh{(\beta E_{\boldsymbol{k}m}/2)}}{E_{\boldsymbol{k}m}}$.

The mean-field parameters $\mu_b$, $\Delta^{b}$, $\mu_a$, $\Delta^a$, $\chi^a$, and $\gamma$ are obtained by solving the saddle-point equations in Eqs.~\eqref{MF equ of b} and \eqref{MF equ of a}. The parameters used in the main text are listed in Table~\ref{tab:typical MF values}.
\begin{table}[h]
\caption{\label{tab:typical MF values}
Self-consistent mean-field parameters used as input for the spectral calculations. We set $t=2J$, and all dimensionful quantities are measured in units of $J$.
}
\begin{ruledtabular}
\begin{tabular}{ccccccc}
Doping ($\delta$) & $\mu_b(J)$ & $\Delta^{b}$ & $\mu_a(J)$ & $\Delta^a$ & $\chi^a$ & $\gamma(J)$\\
\colrule
0.06 & 1.90 & 1.13 & 5.52 & 0.054 & 0.041 & 0.97\\
0.14 & 1.35 & 1.10 & 5.50 & 0.125 & 0.090 & 1.55\\
0.20 & 0.85 & 1.08 & 5.70 & 0.179 & 0.121 & 2.32\\
\end{tabular}
\end{ruledtabular}
\end{table}

\section{A Simple Clarification of the Sign of $t^\prime$ Relating the Electron-Doped and Hole-Doped Sides}
\label{sign of t'}
The high-energy model is described by a doped Hubbard Hamiltonian with nearest- and next-nearest-neighbor hopping. To account for the strong on-site repulsion, the effective Hamiltonian in the reduced Hilbert space is obtained by projecting out the high-energy doubly occupied states, following the conventional canonical transformation~\cite{Auerbach1994}.

It is therefore crucial to identify the appropriate low-energy subspace for different doping regimes. On the hole-doped side, the retained states comprise the empty and singly occupied lattice sites, making a description in terms of physical electron operators natural. In contrast, on the electron-doped side, carrier addition on top of the half-filled background introduces doublons. Here, the low-energy manifold consists of singly occupied states and these doublons, whereas empty sites are highly disfavored. To formally map this electron-doped system onto a mathematically equivalent lower-half Hilbert space, where charge carriers can be treated as vacancy-like excitations, it is mathematically advantageous to adopt a hole-like language, wherein the original physical empty states are cast as doubly occupied hole states that suffer from the high-energy Coulomb penalty.

To explicitly implement this shift in the low-energy description, one performs the particle–hole transformation $c_{i\sigma} \rightarrow (-1)^i c^\dagger_{i\sigma}$. Under this mapping, the nearest-neighbor hopping remains invariant, while the next-nearest-neighbor hopping term changes sign, thereby naturally generating the effective $t$–$t^\prime$–$J$ model with an inverted sign of $t^\prime$.

\section{Calculation of the $a$-Spinon Spectrum Involving $t'$}
\label{a spinon MF with t'}
\begin{figure}[h]
\centering\includegraphics[width=0.6\linewidth]{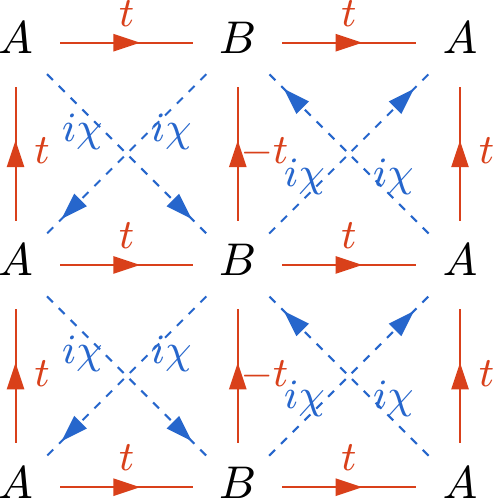}
\caption{Sketch of the $a$-spinon hopping paths in the presence of a background $\pi$ flux. The red and blue arrows denote different gauge choices for the NN and NNN hopping channels.}\label{a spinon hopping sketch}
\end{figure}
\begin{figure}[h]
\centering\includegraphics[width=0.9\linewidth]{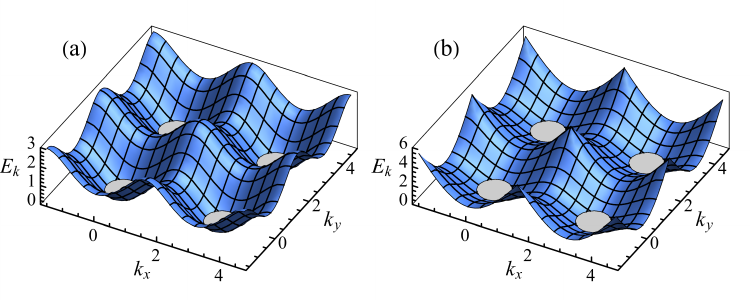}
\caption{The lower branch of the $a$-spinon energy spectrum for (a) $\kappa = 0$ and (b) $\kappa = 0.3t$. The minimum of the dispersion remains invariant under changes in the sign of $\kappa$.}
\label{a spinon spectrum tt'}
\end{figure}

In the phase-string fractionalization scheme, an $a$-spinon experiences a uniform $\pi$ flux per plaquette. The nearest-neighbor hopping Hamiltonian is given by
\begin{equation}
\begin{split}
H_t &\rightarrow -t\sum_{\braket{ij}\sigma}h^\dagger_i h_j e^{iA^s_{ij}}a^\dagger_{i\sigma}a_{j\sigma}e^{-i\phi_{ij}}+\text{H.c.}\\
&\approx 
-t_a\sum_{\braket{ij}\sigma}a^\dagger_{i\sigma}a_{j\sigma}e^{-i\phi_{ij}}+\text{H.c.},
\end{split}
\end{equation}
where $\phi_{ij}$ represents the gauge field associated with the background $\pi$ flux. For the next-nearest-neighbor hopping channel, we have
\begin{equation}
H_{t'} \rightarrow
t'\sum_{\braket{\braket{ij}}\sigma} h^\dagger_i h_j e^{iA^s_{ij}} a^\dagger_{i\sigma} a_{j\sigma} e^{-i\phi_{ij} + i\frac{\sigma-1}{2}\pi} + \text{H.c.}
\end{equation}
Crucially, the phase factor $e^{i\frac{\sigma-1}{2}\pi}$ contributes an integer multiple of $2\pi$ along any closed next-nearest-neighbor hopping loop, and thus can be completely gauged away. Assuming that the holons form a coherent condensate, we can define the gauge-invariant effective next-nearest-neighbor hopping parameter as $\kappa = t'\braket{h^\dagger_i h_j e^{iA^s_{ij}}}$. The background $\pi$ flux per plaquette doubles the unit cell in the horizontal direction, separating the lattice into $A$ and $B$ sublattices as illustrated in Fig.~\ref{a spinon hopping sketch}. To diagonalize the Hamiltonian, we perform the Fourier transformation:
\begin{equation}
\begin{split}
a_{A,\boldsymbol{r}_i,\sigma}&=\frac{1}{\sqrt{N}}\sum_{\boldsymbol{k}}e^{i\boldsymbol{k}\cdot \boldsymbol{r}_i}a_{A,\boldsymbol{k},\sigma},\\
a_{B,\boldsymbol{r}_i+\boldsymbol{a},\sigma}&=\frac{1}{\sqrt{N}}\sum_{\boldsymbol{k}}e^{i\boldsymbol{k}\cdot (\boldsymbol{r}_i+\boldsymbol{a})}a_{B,\boldsymbol{k},\sigma}.\\
\end{split}
\end{equation}
where $\boldsymbol{r}_i$ runs over the reduced magnetic unit cells, the momentum $\boldsymbol{k}$ is restricted to the magnetic Brillouin zone, and $\boldsymbol{a} = (1,0)$ represents the lattice vector. The total effective $a$-spinon Hamiltonian, including the chemical potential term $\mu_a$, is given by $H^a_{t-t'} = H_t + H_{t'} +\mu_a(\sum_{\alpha,\boldsymbol{k},\sigma} a^\dagger_{\alpha,\boldsymbol{k},\sigma} a_{\alpha,\boldsymbol{k},\sigma}-\delta N)$. In the Nambu-like basis, it reads
\begin{equation}
H^a_{t-t'} = \sum_{\boldsymbol{k},\sigma} \psi^\dagger_{\boldsymbol{k},\sigma} h_{\boldsymbol{k}} \psi_{\boldsymbol{k},\sigma} + \text{const.},
\end{equation}
where the spinor is defined as $\psi_{\boldsymbol{k},\sigma} = \begin{pmatrix} a_{A,\boldsymbol{k},\sigma} & a_{B,\boldsymbol{k},\sigma} \end{pmatrix}^T$, and the kernel matrix $h_{\boldsymbol{k}}$ is expressed as
\begin{equation}
\begin{split}
h_{\boldsymbol{k}} &= -2t_a\cos k_y \rho_z - 2t_a\cos k_x \rho_x \\
&+ 4\kappa \sin k_x \sin k_y \rho_y + \mu_a I,
\end{split}
\end{equation}
By diagonalizing $h_{\boldsymbol{k}}$, the energy spectrum is explicitly derived as
\begin{equation}
E_{\boldsymbol{k}} = \pm\sqrt{4t_a^2(\cos^2 k_x + \cos^2 k_y) + 16\kappa^2 \sin^2 k_x \sin^2 k_y} + \mu_a.
\end{equation}
Notably, because the energy spectrum $E_{\boldsymbol{k}}$ depends exclusively on $\kappa^2 \propto t'^2$, we conclude that the sign of $t'$ has no impact on the positions of the pocket minima. The lower branch of the resulting $a$-spinon dispersion is shown in Fig.~\ref{a spinon spectrum tt'} for distinct ratios of $\kappa/t$.

\section{VMC and DMRG Calculations on the Single-Charge-Doped System}
\label{numerical_results}
The calculated ground-state energies of the \emph{Ansatz} in Eq.~\eqref{singlehole} on an $8\times 8$ lattice with $t' = \pm0.25t$ are listed in Table~\ref{tab:table1}. For comparison, we also perform DMRG calculations \cite{10.21468/SciPostPhysCodeb.4-r0.3} as a benchmark, keeping bond dimensions up to $D = 5000$ to ensure accurate results with a maximum truncation error $\epsilon \sim 4\times 10^{-6}$. The spin-$U(1)$ symmetry is imposed by fixing the total $S^z = -\frac{1}{2}$ without loss of generality. The VMC ground-state energies are found to be in close quantitative agreement with the DMRG results. 

\begin{table}[h]
\caption{\label{tab:table1}%
Single-hole ground-state energies (in units of $J$) calculated using VMC and DMRG methods on an $8 \times 8$ lattice. $E_{t}$, $E_{t'}$, and $E_J$ denote the kinetic energies from the NN and NNN hopping channels and the superexchange energy, respectively. The NN hopping parameter is $t = 2J$.
}
\begin{ruledtabular}
\begin{tabular}{ccccc}
&
\textrm{$E_{t}$}&
\textrm{$E_{t'}$}&
\textrm{$E_J$}&
\textrm{$E_{tot}$}\\
\colrule

$|\Psi_{G}\rangle_{1\mathrm{h}}(t'=-0.25t)$ & -4.23 & -0.31 & -65.05 & -69.60\\
DMRG$(t'=-0.25t)$ & -5.52  & -0.51 & -64.53 & -70.56\\
$|\Psi_{G}\rangle_{1\mathrm{h}}(t'=0.25t)$ & -3.88 & -0.89 & -64.72 & -69.50\\
DMRG$(t'=0.25t)$ & -4.82  & -0.78 & -64.66 & -70.26\\
\end{tabular}
\end{ruledtabular}
\end{table}

To probe the quasiparticle component $c_{\boldsymbol{k}\uparrow}$ in the low-energy regime, we calculate the quasiparticle spectral weight of the variational ground state $|\Psi_{1\mathrm{h}}\rangle$:
\begin{equation}
Z_{\boldsymbol{k}}=|\langle\phi_0|c^{\dagger}_{\boldsymbol{k}\uparrow}|\Psi_{1\mathrm{h}}\rangle|^2
\label{specweight_def}
\end{equation}

\begin{figure}[h]
\centering
\includegraphics[width=0.9\linewidth]{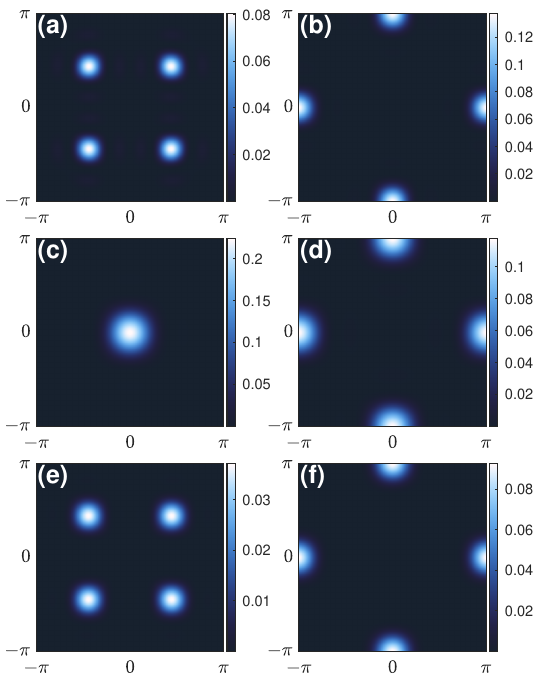}
\caption{Quasiparticle spectral weight $Z_{\boldsymbol{k}}$ (defined in Eq.~\eqref{specweight_def}) computed for the single-hole ground state $|\Psi_{1\mathrm{h}}\rangle$ using VMC and DMRG methods. (a),(b) $|\Psi_{1\mathrm{h}}\rangle = |\Psi_G\rangle_{1\mathrm{h}}$ for $t'=-0.25t$ and $t'=+0.25t$, respectively. (c),(d) $|\Psi_{1\mathrm{h}}\rangle = |\Psi_{\text{Bloch}}\rangle_{1\mathrm{h}}$ for $t'=-0.25t$ and $t'=+0.25t$, respectively. (e),(f) DMRG results for $t'=-0.25t$ and $t'=+0.25t$, respectively.}
\label{zkVMC}
\end{figure}

For $|\Psi_{1\mathrm{h}}\rangle = |\Psi_G\rangle_{1\mathrm{h}}$, the calculated $Z_{\boldsymbol{k}}$ for $t'= \pm 0.25t$ is shown in Figs.~\ref{zkVMC}(a) and \ref{zkVMC}(b), respectively.  In the hole-doped case ($t' = -0.25t$), $Z_{\boldsymbol{k}}$ peaks at $(\pm \pi/2, \pm \pi/2)$, whereas in the electron-doped case ($t' = 0.25t$), it peaks at $(\pi, 0)$ and $(0, \pi)$, in agreement with the DMRG results in Figs.~\ref{zkVMC}(e) and \ref{zkVMC}(f).

For the Bloch-wave ansatz $|\Psi_{1\mathrm{h}}\rangle = |\Psi_{\text{Bloch}}\rangle_{1\mathrm{h}}$, shown in Figs.~\ref{zkVMC}(c) and \ref{zkVMC}(d), $Z_{\boldsymbol{k}}$ incorrectly peaks at $(0, 0)$ for $t'= -0.25t$, but correctly peaks at $(\pi, 0)$ and $(0,\pi)$ for $t' = 0.25t$. This contrast highlights the qualitatively different nature of the low-energy carriers in the hole- and electron-doped regimes, as discussed in the main text.


\section{Simplification of $e^{-\frac{i}{2}(\Phi_i^0-\Phi_j^0)}$ in Eq.~\eqref{D and f}}
\label{simplification of phase factor}
By inserting a sequence of nearest-neighbor links from $i$ to $j$, denoted by $i_1$, $i_2$, ..., $i_M$, we have
\begin{equation}
\begin{split}
    &\exp\{-\frac{i}{2}(\Phi_i^0-\Phi_j^0)\}\\
    =&\exp\{-\frac{i}{2}(\Phi_i^0-\Phi_{i_1}^0+\Phi_{i_1}^0-\Phi_{i_2}^0+\cdots+\Phi_{i_M}^0-\Phi_j^0)\}\\
    =&\exp(-i\sum_{i\rightarrow j}\phi_{i_s,i_{s+1}}^0)\prod_{i\rightarrow j}\exp
    \{-\frac{i}{2}(\theta_{i_s}(i_{s+1})-\theta_{i_{s+1}}(i_s))\},
\end{split}
\end{equation}
where $\phi_{i_{s},i_{s+1}}^0=\frac{1}{2}\sum_{l\neq i_{s},i_{s+1}}[\theta_{i_s}(l)-\theta_{i_{s+1}}(l)]$. Since $\theta_{i_s}(i_{s+1})-\theta_{i_{s+1}}(i_{s})=\pm \pi$, the product term can be further simplified as 
\begin{equation}
\begin{split}
    &\prod_{i\rightarrow j}\exp
    \{-\frac{i}{2}[\theta_{i_s}(i_{s+1})-\theta_{i_{s+1}}(i_s)]\}\\
    =&(e^{\pm i\frac{\pi}{2}})^{(i-j)}=e^{i \boldsymbol{k}_0\cdot (\boldsymbol{r}_i-\boldsymbol{r}_j)},
\end{split}
\end{equation}
$\boldsymbol{k}_0=(\pm\pi/2,\pm\pi/2)$. We choose the sign of the angle difference to be the same on each link to keep the phase factor smooth. Finally, we have
\begin{equation}
    \exp\{-\frac{i}{2}(\Phi_i^0-\Phi_j^0)\}=\exp(-i\sum_{i\rightarrow j}\phi_{i_s,i_{s+1}}^0)e^{i \boldsymbol{k}_0\cdot (\boldsymbol{r}_i-\boldsymbol{r}_j)}.
\end{equation}
The factor $\exp(-i\sum_{i\rightarrow j}\phi_{i_s,i_{s+1}}^0)$ is the phase factor accumulated due to the background $\pi$ flux experienced by $\tilde{c}$ and should be absorbed into the propagator of $\tilde{c}$ to make it gauge invariant, similar to the usual Peierls substitution.

\section{Derivation of $D^{\tilde{c}}$}
\label{Self energy correction}
The detailed correction to $D^{\tilde{c}}$, which is used in Sec.~\ref{Precise calculation of single particle spectrum}, is introduced in this section.

In the mean-field calculation in Appendix~\ref{Effective Hamiltonian}, $\tilde{c}_{i\sigma}=h_i^\dagger a_{i\bar\sigma}^\dagger$. Since $\braket{h_i^\dagger}\neq 0$, the propagator of $\tilde{c}$ follows that of $a^\dagger$. As a result, the generic bare propagator of $\tilde{c}$ is expressed as a matrix, denoted by $\hat{D}^{\tilde{c}}_0$, in the Nambu spinor basis:
\begin{equation}
\hat{D}_0^{\tilde{c}}(\boldsymbol{r}_{i}-\boldsymbol{r}_{j},\tau)
=-\left\langle
\begin{pmatrix}
\tilde{c}_{i\uparrow}(\tau)\\
\tilde{c}_{i\downarrow}^{\dagger}(\tau)
\end{pmatrix}
\begin{pmatrix}
\tilde{c}_{j\uparrow}^{\dagger}(0) &\tilde{c}_{j\downarrow}(0)
\end{pmatrix}\right\rangle,
\label{real_space_D0}
\end{equation}
where the circumflex hat denotes a $2 \times 2$ matrix in Nambu space. Transforming this expression to momentum and frequency space, we have
\begin{widetext}
\begin{equation}
    \hat{D}^{\tilde{c}}_0(\boldsymbol{k},i\omega_n)
    =\begin{pmatrix}
        G^{\tilde{c}}_{0\uparrow \uparrow}(\boldsymbol{k},i\omega_n) & F_{0}^{\tilde{c}}(\boldsymbol{k},i\omega_n)\\
        F^{\tilde{c}*}_{0}(\boldsymbol{k},i\omega_n) & -G^{\tilde{c}}_{0\downarrow \downarrow}(-\boldsymbol{k}, -i\omega_n)
    \end{pmatrix}
    =\frac{1}{i\omega_n\rho_0+\xi_{\boldsymbol{k}}\rho_z-\Delta_{\boldsymbol{k}}\rho_x}
    \label{Nambu}
\end{equation}
\end{widetext}
In the derivation of the Green's function in Sec.~\ref{PhenoGreenFunc}, we use only the normal Green's function,
\begin{equation}
D^{\tilde{c}}_{0}(\boldsymbol{k},i\omega_n)=\diagram{Figures/Gctilde0.pdf}{8pt}=G^{\tilde{c}}_{\uparrow \uparrow}(\boldsymbol{k},i\omega_n).
\end{equation}
Since $c$ and $\tilde{c}$ can be converted into each other by combining with or separating from the vortex operator $e^{i\hat{\Omega}}$, we should introduce the interaction vertex
\begin{equation}
\lambda 
\begin{pmatrix}
    \tilde{c}^\dagger_{i\uparrow}(\tau) &
    \tilde{c}_{i\downarrow}(\tau)
\end{pmatrix}
\begin{pmatrix}
    c_{i\uparrow}(\tau) e^{i\hat{\Omega}_i(\tau)} \\
    c^\dagger_{i\downarrow}(\tau) e^{-i\hat{\Omega}_i(\tau)}
\end{pmatrix}.
\label{vertex}
\end{equation}
Combining Eq.~\eqref{Nambu} and Eq.~\eqref{vertex}, the dressed Green's function $\hat{D}^{\tilde{c}}(\boldsymbol{k},i\omega_n)$ can be written in the following form:
\begin{widetext}
\begin{equation}
    \hat{D}^{\tilde{c}}(\boldsymbol{r}_i-\boldsymbol{r}_j,\tau_i-\tau_j)
    =\hat{D}_0^{\tilde{c}}(\boldsymbol{r}_i-\boldsymbol{r}_j,\tau_i-\tau_j)\\
    +\lambda^2 \sum_{j', j''} \int \mathrm{d}\tau' \mathrm{d}\tau'' \, 
    \hat{D}_0^{\tilde{c}}(\boldsymbol{r}_i - \boldsymbol{r}_{j'}, \tau_i - \tau') 
    \hat{\Sigma}^{\tilde{c}}(\boldsymbol{r}_{j'} - \boldsymbol{r}_{j''}, \tau' - \tau'') 
    \hat{D}_0^{\tilde{c}}(\boldsymbol{r}_{j''} - \boldsymbol{r}_j, \tau'' - \tau_j)+...
\end{equation} 
\end{widetext}

\begin{widetext}
\begin{equation}
\hat\Sigma^{\tilde{c}}(\boldsymbol{r}_{i}-\boldsymbol{r}_{j},\tau)=-\lambda^2\left\langle
\begin{pmatrix}
c_{i\uparrow}(\tau)e^{-i\hat{\Omega}_{i}(\tau)} \\
c_{i\downarrow}^{\dagger}(\tau)e^{i\hat{\Omega}_{i}(\tau)}
\end{pmatrix}
\begin{pmatrix}
c_{j\uparrow}^{\dagger}(0)e^{i\hat{\Omega}_{j}(0)}&c_{j\downarrow}(0)e^{-i\hat{\Omega}_{j}(0)}
\end{pmatrix}\right\rangle,
\label{self-energy}
\end{equation}
\end{widetext}
which has been discussed in Ref.~\cite{ZhangJianHao}. In explicit form, the Dyson equation for the dressed propagator of $\tilde{c}$ at the RPA level reads
\begin{equation}
\hat{D}^{\tilde{c}}(\boldsymbol{k},i\omega_n) = \frac{1}{[\hat{D}_0^{\tilde{c}}(\boldsymbol{k},i\omega_n)]^{-1} - \hat\Sigma^{\tilde{c}}(\boldsymbol{k},i\omega_n)}.
\label{Dctilde}
\end{equation}

\begin{figure}
   \centering
   \includegraphics[width=0.8\linewidth]{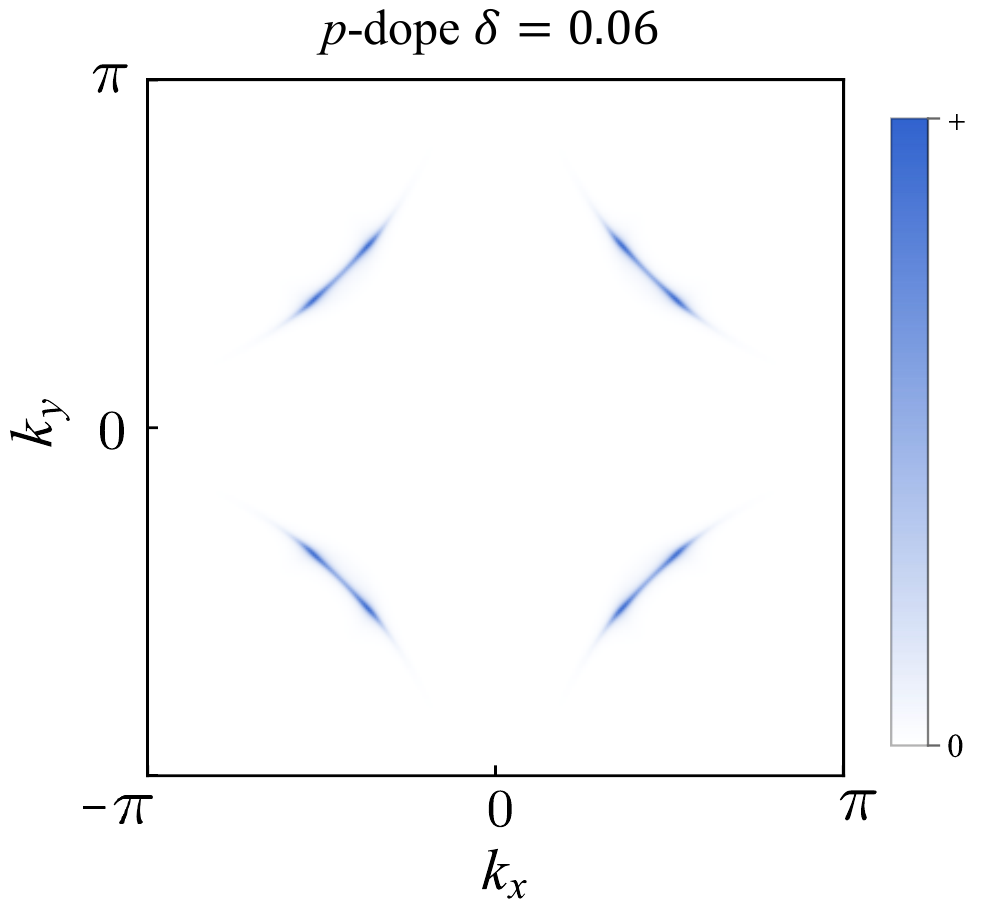}
   \caption{$-\mathrm{Im}\ \tilde{G}_{0}^{e}(\boldsymbol{k},0)$ at $\delta=0.06$, with spectral weight centered at $(\pm \pi/2, \pm\pi/2)$. Parameters corresponding to the $\delta=0.06$ row of Table~\ref{tab:typical MF values} are used, with $\lambda=0.06J$.}
   \label{Actilde}
\end{figure}

In the lower pseudogap phase characterized by condensed holons and deconfined $b$-spinons~\cite{Crossover_Fermi_arc,Weng_Qi_LPP}, long-range phase coherence implies that the propagator of the phase factor $e^{i\hat{\Omega}_i(\tau)}$ saturates at long distances and long times.  $$\left\langle e^{i\left[\hat{\Omega}_i(\tau)-\hat{\Omega}_j(0)\right]}\right\rangle=e^{ik_0(\boldsymbol{r}_i-\boldsymbol{r}_j)}f_0(\boldsymbol{r}_i-\boldsymbol{r}_j,\tau),$$
while the anomalous propagator of the phase factor reads $$\left\langle e^{i\left[\hat{\Omega}_i(\tau)+\hat{\Omega}_j(0)\right]}\right\rangle=\left\langle e^{2i\hat{\Omega}_j(0)}\right\rangle e^{ik_0(\boldsymbol{r}_i-\boldsymbol{r}_j)}f_0(\boldsymbol{r}_i-\boldsymbol{r}_j,\tau)=0.$$
Therefore, the anomalous term in $\hat{\Sigma}^{\tilde{c}}$ vanishes. Beyond the AFM correlation length, $f_0(\boldsymbol{r}_i-\boldsymbol{r}_j,\tau)$ approaches a constant $F_0$, whose precise value is unimportant and is chosen to be 1 throughout this paper. Consequently, $\hat{\Sigma}^{\tilde{c}}$ is directly related to $G_0^c$, the non-interacting propagator of $c$ expressed in Eq.~\eqref{Gc0}, in the following sense:
\begin{equation}
\begin{split}
&\hat{\Sigma}^{\tilde{c}}(\boldsymbol{k},i\omega_n)=\hat{G}^c_0(\boldsymbol{k}-\boldsymbol{k}_0,i\omega_n)F_0 \\
&= 
\left(
    \begin{array}{cc}
        G^{c}_{0\uparrow \uparrow}(\boldsymbol{k}-\boldsymbol{k}_0,i\omega_n) & \\
        & -G^{c}_{0\downarrow \downarrow}(-\boldsymbol{k}+\boldsymbol{k}_0 ,-i\omega_n)
    \end{array}
\right)F_0.
\end{split}
\label{another-self-energy}
\end{equation}
The dressed $\tilde{c}$ propagator, $[\hat{D}^{\tilde{c}}(\boldsymbol{k},i\omega_n)]_{11}$, can be derived by combining Eq.~\eqref{Nambu} with Eq.~\eqref{another-self-energy}. For the calculations in the main text, a more relevant Green's function is $\tilde{G}_0^{e}(\boldsymbol{k},0)=[\hat{D}^{\tilde{c}}(\boldsymbol{k}-\boldsymbol{k}_0,0)]_{11}F_0$, whose spectral weight is displayed in Fig.~\ref{Actilde}.

\section{Approach Using the Nonlinear $\sigma$ Model}
\label{App NLSM}
While the main text incorporates the effect of $\ket{\Phi_b}$ into the AFM folding calculation by assuming a simplified, static local-moment background, this section introduces a more rigorous dynamical treatment.In the absence of long-range AFM order, the dynamics of short-range spin fluctuations are explicitly captured by the O(3) nonlinear $\sigma$ model (NL$\sigma$M), governed by the action~\cite{Auerbach1994}:
\begin{equation}
\begin{split}
S = \frac{1}{2g} \int d\tau d^2\boldsymbol{r} \Big[ &(\partial_\tau \boldsymbol{n}^b)^2 + v_s^2 (\nabla \boldsymbol{n}^b)^2 \\
&+ i\lambda(\boldsymbol{r}, \tau) \left( (\boldsymbol{n}^b)^2 - 1 \right) \Big].
\end{split}
\end{equation}
Here, the mean-field expectation value of the Lagrange multiplier defines the spin gap, $\langle i\lambda \rangle = m_s^2 > 0$, where $m_s$ is inversely proportional to the spin correlation length ($\xi_s^{-1}$). By relaxing the rigid length constraint on the N'eel field $\boldsymbol{n}^b$ to accommodate this finite correlation length, the effective action in imaginary time $\tau$ reduces to a Gaussian form:
\begin{equation}
S = \frac{1}{2g} \int d\tau d^2\boldsymbol{r} \left[ (\partial_\tau \boldsymbol{n}^b)^2 + v_s^2 (\nabla \boldsymbol{n}^b)^2 + m_s^2(\boldsymbol{n}^b)^2 \right],
\end{equation}
where $\boldsymbol{n}^b(\boldsymbol{r},\tau)$ denotes the fluctuating staggered magnetization field, $v_s$ is the spin-wave velocity, and the mass term $m_s$ characterizes the spin gap that dictates the short-range AFM correlations.

From this Gaussian effective action, we can directly extract the bare propagator for the spin fluctuations. In momentum and Matsubara frequency space, the low-energy dispersion relation of the spin excitations is given by:
\begin{equation}
E^n_{\boldsymbol{k}}=\sqrt{v_s^2\boldsymbol{k}^2+m_s^2},
\end{equation}
and the corresponding dynamic spin susceptibility is expressed as:
\begin{equation}
\begin{split}
    \chi^{b}(\boldsymbol{k},i\omega_n)&= \braket{\boldsymbol{n}^b(\boldsymbol{k},i\omega_n)\cdot\boldsymbol{n}^b(-\boldsymbol{k},-i\omega_n)}\\
    &= -\frac{g}{(i\omega_n)^2-(E^n_{\boldsymbol{k}})^2}.
\end{split}
\end{equation}
These dynamical spin fluctuations couple to the physical electrons via the effective interaction Hamiltonian:
\begin{equation}
H_{\text{int}}=J_{\text{cp}}\sum_i e^{i \boldsymbol{k}_{\text{AF}}\cdot \boldsymbol{r}_i} \boldsymbol{n}_i^b\cdot c_{i\alpha}^\dagger \boldsymbol{\rho}_{\alpha\beta} c_{i\beta},
\end{equation}
where $\boldsymbol{k}_{\text{AF}}=(\pi,\pi)$ is the AFM nesting vector and $J_{\text{cp}}$ represents the coupling strength.

Formally, the primary departure from the approach in the main text is the replacement of the static AFM order assumed in Eq.~\eqref{folded equation} with these dynamical, short-range AFM fluctuations. Nevertheless, the core physical mechanism remains intact: the virtual emission and absorption of these spin fluctuations still mediate strong electron scattering near the AFM Brillouin zone boundary. Ultimately, this dynamical scattering yields a spectral weight suppression that closely mirrors the static folding effect.

\end{document}